\documentclass[superscriptaddress, nofootinbib, amsfonts, amsmath, amssymb, aps, prd, notitlepage, floatfix]{revtex4-2}
\usepackage[colorlinks=true, linkcolor=blue, citecolor=blue, urlcolor=blue]{hyperref}
\usepackage{xcolor}
\usepackage{graphicx}
\usepackage{subfigure}
\usepackage{multirow}
\usepackage{slashed}
\usepackage{lineno}
\usepackage[percent]{overpic}
\usepackage{ulem}
\allowdisplaybreaks

\newcommand{\na}{\notag\\} 
\newcommand{\bb}{\mathbb}
\newcommand{\rP}{\rm P}
\newcommand{\dhd}{{\textstyle d} \lower.03ex\hbox{\kern-0.38em$^{\scriptstyle-}$}\kern-0.05em{}}
\usepackage{xcolor}
\definecolor{myred}{RGB}{188,42,30}

\usepackage[defaultcolor=orange]{changes}

\begin{document}
%
\title{Back-to-back dijet production in DIS at arbitrary Bjorken-x:\\ TMD quark distributions to twist-3 accuracy}
%
%
\author{Swagato~Mukherjee}
\affiliation{Physics Department, Brookhaven National Laboratory, Upton, New York 11973, USA}
\author{Vladimir~V.~Skokov}
\affiliation{Department of Physics and Astronomy, North Carolina State University, Raleigh, NC 27695, USA}
\author{Andrey~Tarasov}
\affiliation{Department of Physics and Astronomy, North Carolina State University, Raleigh, NC 27695, USA}
\affiliation{Center for Frontiers in Nuclear Science (CFNS) at Stony Brook University, Stony Brook, NY 11794, USA}
\author{Shaswat~Tiwari}
\email{sstiwari@ncsu.edu}
\affiliation{Physics Department, Brookhaven National Laboratory, Upton, New York 11973, USA}
\affiliation{Department of Physics and Astronomy, North Carolina State University, Raleigh, NC 27695, USA}
\author{Fei~Yao}
\email{fyao@bnl.gov}
\affiliation{Physics Department, Brookhaven National Laboratory, Upton, New York 11973, USA}
\begin{abstract}
{We study the contributions of target quark background to quark-gluon and quark-antiquark dijet productions cross sections in deep inelastic scattering (DIS) at arbitrary Bjorken-$x$. Using the background-field method, we organize the leading-order dijet production cross section in terms of bilocal quark and trilocal quark--gluon--quark correlators. The former provides the contributions of the leading-twist as well as the kinematical and dynamical twist-three quark transverse-momentum-dependent (TMD) distributions, while the latter encode contributions of the dynamical twist-three TMD distributions.  {Together with the gluon distributions derived in Ref.~\cite{Mukherjee:2026cte}}, these results provide the complete leading-order cross section for back-to-back dijet production in DIS at arbitrary Bjorken-$x$ to twist-three accuracy. This theoretical framework extends the TMD description of the process beyond the small-$x$ eikonal regime and provides the necessary tools for analyzing data from the future Electron--Ion Collider.}
\end{abstract}

\date{\today}
\maketitle


\section{Introduction}

Understanding the transverse-momentum-dependent (TMD) partonic structure of nucleons and nuclei is one of the central goals of the future Electron--Ion Collider (EIC)~\cite{Accardi:2012qut,AbdulKhalek:2021gbh}. TMD distributions encode the dependence of parton densities on the longitudinal momentum fraction and intrinsic transverse momentum, providing a three-dimensional picture of the target in momentum space \cite{Collins:2011zzd,Ji:2004wu,Bacchetta:2006tn,Boussarie:2023izj}. These distributions can be accessed through TMD factorization {of} experimentally measured cross sections of {processes, such as the} semi-inclusive deep inelastic scattering (SIDIS), Drell-Yan and dijet production in DIS in appropriate kinematic limits. Among these processes, dijet production in DIS  {in near back-to-back kinematics} provides a particularly useful probe of TMD dynamics. 

{Dijet} production in DIS has emerged as one of the cleanest probes of the transverse structure of the target, especially of the Weizsaecker--Williams gluon distribution at small $x$
\cite{Dominguez:2011wm,Dominguez:2011br,Metz:2011wb}. In the correlation (back-to-back) limit, where the transverse-momentum imbalance of the two jets is much smaller than their individual transverse momenta, the hierarchy of scales
makes the transverse momentum of the target partons experimentally accessible. The cross section can then be factorized into perturbatively calculable hard coefficients and nonperturbative TMD distributions. In a previous work~\cite{Mukherjee:2026cte}, we {studied quark-antiquark} dijet production in {target gluon} background and showed the factorization of dijet cross section in terms of gluon TMD operators at arbitrary Bjorken-$x$ to twist-three accuracy. 

The factorization of dijet production has typically been studied using two complementary approaches. At large $x$, soft-collinear effective theory and standard TMD factorization \cite{Collins:2011zzd,Mulders:2000sh,delCastillo:2020omr} describe the process at leading twist, including next-to-leading order (NLO) corrections in the strong coupling. At small $x$, the Color Glass Condensate (CGC) effective theory
\cite{McLerran:1993ni,McLerran:1994vd,Iancu:2003xm,Gelis:2010nm} was used in the eikonal, high-energy limit to factorize the dijet production cross section in a gluon background, including NLO corrections and twist-three contributions
\cite{Dominguez:2011wm,Boussarie:2021ybe,Caucal:2021ent,Caucal:2023fsf, Caucal:2023nci,Altinoluk:2022jkk,Altinoluk:2024zom}. At sub-eikonal accuracy, the CGC framework also gives access to quark TMDs at small $x$ through quark-exchange channels, although these calculations do not retain the finite-$x$ longitudinal phase \cite{Altinoluk:2023qfr}. A factorization of the dijet production cross section in a quark background
field at arbitrary Bjorken-$x$ and to twist-three accuracy has therefore remained unavailable.

Apart from completing the twist-three description of dijet production, there are also practical reasons to include quark contributions at EIC kinematics. Quark- and gluon-initiated jets cannot be perfectly distinguished on an event-by-event basis at the EIC~\cite{Singh:2026wol,Gras:2017jty,Frye:2017yrw}.
Therefore, both the quark--antiquark final state studied in Ref.~\cite{Mukherjee:2026cte} and the quark--gluon final state studied in the present work are needed for a complete description of dijet events. Moreover, sub-eikonal effects can become numerically important at the accessible values $x \gtrsim 10^{-2}$ relevant for EIC kinematics~\cite{Dumitru:2018kuw}. A consistent treatment at twist-three accuracy should therefore include the operators constructed from background quark fields.
  \par

Our analysis is based on the background-field operator approach developed in Refs.~\cite{Mukherjee:2023snp, Mukherjee:2025aiw}, hereafter referred to as the MSTT framework. It retains the full longitudinal phase $e^{i x P^+ z^-}$ and is therefore valid at arbitrary Bjorken-$x$. We organize the tree-level amplitudes of the quark--gluon and quark--antiquark channels in terms of background-field propagators, expand them in the back-to-back limit, and reduce the squared amplitude to a gauge-invariant basis of bilocal and trilocal quark operators. We present the explicit hard coefficients for longitudinally and transversely polarized virtual photons, and we verify that the leading-twist terms reproduce the known results of Ref.~\cite{Altinoluk:2023qfr}. These results extend the TMD description of dijet production beyond the small-$x$ eikonal regime and provide ingredients needed for precision studies of the quark content of the nucleon at the EIC. \par

The paper is organized as follows. In Sec.~\ref{quark_bg:amp} we set up the kinematics and the background-field notation and present the general structure of the tree-level dijet amplitudes in a quark background for both the quark–gluon and quark–antiquark channels. In Sec.~\ref{propagators}, we organize the background-field propagators needed in this work. These include the LSZ-reduced external quark and gluon legs, as well as the internal quark and gluon propagators in a quark background, expanded to the single-field-strength accuracy required at twist three. Section~\ref{sec:dijet-amplitudes} uses these building blocks to construct the explicit dijet amplitudes for the two quark–gluon topologies and for the quark–antiquark channel, expanded to the accuracy needed in the back-to-back limit. Section~\ref{sec:cross_section_operator_structure} reorganizes the results at the level of the cross section, separating the bilocal and trilocal operator contributions and giving the longitudinal and transverse hard coefficients to twist-three accuracy. We summarize our results and discuss their implications in Sec.~\ref{sec:conclusion}.

\section{General structure of the dijet amplitude and cross section}
\label{quark_bg:amp}
 
In Ref.~\cite{Mukherjee:2026cte}, we derived the cross section for the back-to-back quark-antiquark dijet production in deep inelastic scattering at arbitrary Bjorken-$x$ in the presence of background gluon fields. Here, 
we complete the calculation by including quark background fields. This extension gives access to the quark TMD sector, which receives contributions from both quark-antiquark and quark-gluon dijet channels. 

In this section, we first present the general formulation of the corresponding amplitudes and cross section in quark background, and then discuss the individual channels.

\subsection{General setup} 
In this work, we follow the same kinematic conventions and Schwinger notation as in Ref.~\cite{Mukherjee:2026cte}. To access the quark TMD sector, we consider dijet production in the presence of target quark background fields. Specifically, we study
\begin{equation}
\gamma^*(q)+h(P)\to j_1(k_1)+j_2(k_2)+X,
\qquad
j_1j_2=q\bar q,\ qg .
\label{eq:process_quark_bg}
\end{equation}
Here \(qg\) and \(q\bar q\) label the partonic content of the two measured jets.  In the \(qg\) channel, \(j_1=q\) is the produced quark jet and \(j_2=g\) is the produced gluon jet.  In the \(q\bar q\) channel, \(j_1=q\) and \(j_2=\bar q\) are the produced quark and antiquark jets.  In the quark-background calculation these final states arise from amplitudes with explicit insertions of the target quark fields: the \(qg\) channel contains one insertion of either \(\psi\) or \(\bar\psi\), while the \(q\bar q\) channel contains a pair of insertions, \(\psi\) and \(\bar\psi\).

The {dijet production} amplitudes {naturally involve the} background-field propagators.  We use \(\hat{\mathcal S}\) for the fermion propagator that {goes into the final state (i.e. requires LSZ-amputation)}, and \(\hat{\mathcal S}_{\slashed{\rm P}}\) for internal fermion lines.  The gluon propagator is denoted by \(\hat{\mathcal G}_{\alpha\beta}\); when one of its endpoints is attached to an external gluon, the LSZ reduction is {implied}.  
Explicitly,
\begin{align}
\hat{\mathcal S}
&\equiv
\frac{ 1}{{\rm P^2}+\frac12\sigma F+i\epsilon},
\\
\hat{\mathcal S}_{\slashed{\rm P}}
&\equiv
\slashed{\rm P}\,
\frac{ 1}{{\rm P^2}+\frac12\sigma F+i\epsilon},
\\
\hat{\mathcal G}_{\alpha\beta}
&\equiv
\frac{-i}{g_{\alpha\beta}{\rm P^2}+2iF_{\alpha\beta}+i\epsilon},
\end{align}
where  \(\slashed{\rm P} = \slashed{p} +\slashed{A}\)\footnote{\(A_\mu(x)=t^aA_\mu^a(x)\) is the background gauge field in the fundamental representation. The QCD coupling \(g_s\) is absorbed into \(A_\mu\).}.

The corresponding matrix elements are written as
\begin{align}
\mathcal S(A,B)
&\equiv
 (A |\hat{\mathcal S} |B ),
\\
\mathcal S_{\slashed{\rm P}}(A,B)
&\equiv
 (A |\hat{\mathcal S}_{\slashed{\rm P}} |B ),
\\
\mathcal G_{\alpha\beta}(A,B)
&\equiv
 (A |\hat{\mathcal G}_{\alpha\beta} |B ).
\end{align}
The arguments specify the representation of the  endpoints: they can be coordinate or momentum eigenstates. For example,
\begin{align}
\mathcal S(k,y)
&=
 (k |\hat{\mathcal S} |y )
=
\int d^4x\, e^{ik\cdot x}\,
\mathcal S(x,y),
\\
\mathcal S(x,k)
&=
 (x |\hat{\mathcal S} |k )
=
\int d^4y\,
\mathcal S(x,y)\,
e^{-ik\cdot y},
\end{align}
where we use the convention $(x|k)=e^{-ik\cdot x}$. 

In light-cone variables, the Lorentz-invariant phase-space measure for particle $i$ is
\begin{equation}
\frac{d k_i^-\, d^2 k_{i\perp}}{2 k_i^- (2\pi)^3}.
\label{eq:phase_space_general}
\end{equation}
Accordingly, for a given channel $c$, the differential cross section can be written as
\begin{equation}
k_1^- k_2^-
\frac{d\sigma_c}
{\dhd k_1^- \dhd^2 k_{1\perp}\, \dhd k_2^- \dhd^2 k_{2\perp}}
=
\pi q^- \delta(k_1^-+k_2^- - q^-)\,
|i\mathcal{M}_c|^2 ,
\label{eq:xsec_channel_general}
\end{equation}
where $c=q\bar q$ or $qg$, and factors of $2\pi$ from the phase-space integration {were} absorbed into the definition $\dhd k \equiv dk/(2\pi)$. For later convenience, we define the momentum transferred from the target as
\begin{equation}
k_q \equiv k_1+k_2-q.
\end{equation}
When the amplitudes are expanded below, we use the back-to-back parametrization summarized in Appendix~\ref{app:btb_kinematics}, including the corresponding on-shell expressions for \(k_1\), \(k_2\), and \(k_q\).

The full quark-background initiated contribution is then obtained by summing over the relevant channels,
\begin{equation}
d\sigma 
=
d\sigma_{qg}+d\sigma_{q\bar q}.
\label{eq:xsec_channel_sum}
\end{equation}
In the following subsections, we present the tree-level amplitudes for the \(qg\) and \(q\bar q\) channels. At this stage, the interaction with the target background is kept in the propagator blocks, while the twist expansion of these blocks is carried out in Sec.~\ref{sec:dijet-amplitudes} .

\subsection{Quark-gluon dijet channel}
\begin{figure}[t]
\begin{center}
\begin{overpic}[width=0.7\textwidth]{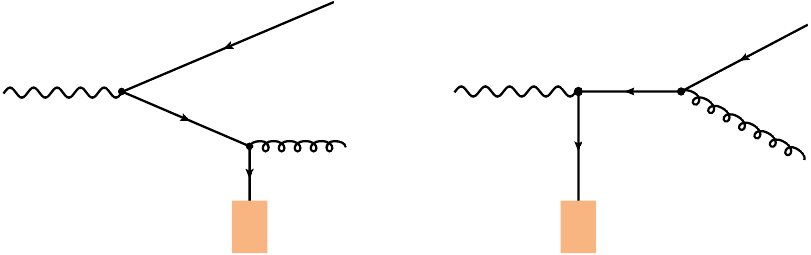}
    \put(3,23){ $ {\gamma^*\,(q)}$} 
    \put(42,31){ $ {k_1}$} 
    \put(43,13){ $ {k_2}$} 
    \put(28.7, 2.8){ $\boldsymbol{\bar\psi }$}
    \put(58,23){ $ {\gamma^*\,(q)}$} 
    \put(100,27.5){ ${k_1}$} 
    \put(100,12.5){ ${k_2}$} 
    \put(69.3, 2.8){ $\boldsymbol{\bar\psi }$}
\end{overpic}
\end{center}
\caption{Tree-level diagrams for quark--gluon dijet production in a quark background. The shaded box denotes the background quark field \(\bar\psi\). The left and right panels correspond to topology \((a)\) and topology \((b)\), respectively, representing the two topologically distinct insertions of the background field. The produced quark and gluon carry momenta $k_1$ and $k_2$, respectively.}
\label{fig:qg_dijet_diagram}
\end{figure}
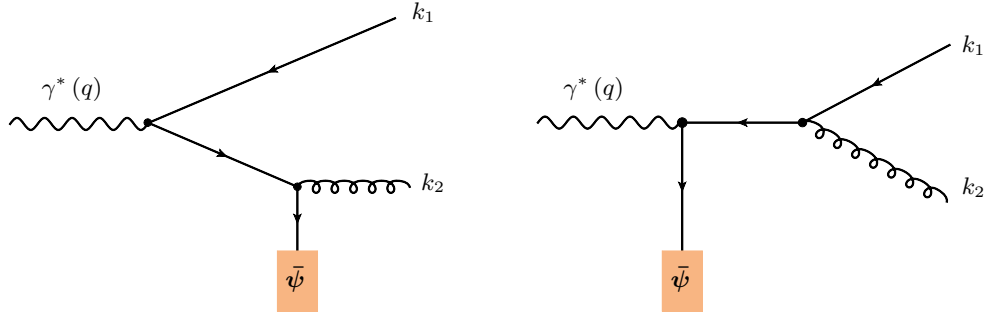
At tree level, quark--gluon dijet production in a quark background receives contributions from the two topologies shown in Fig.~\ref{fig:qg_dijet_diagram}. The produced quark and gluon carry momenta $k_1$ and $k_2$, respectively. For each topology, the \(qg\) channel receives a pair of charge-conjugate contributions, distinguished by whether the explicit target-background insertion is \(\bar\psi\) or \(\psi\). We first consider the topology shown in the left panel of Fig.~\ref{fig:qg_dijet_diagram}, in which the target-background insertion and the final-state gluon emission lie on the same fermion line. The \(\bar\psi\)- and \(\psi\)-insertion contributions can be combined as

\begin{align}
\label{eq:fig1_qg_a}
i\mathcal M_{qg}^{(a)}
=&
-ie\,
\lim_{k_1^2,k_2^2\to0}
k_1^2 k_2^2
\int d^{\,4}x\,d^4y\,
 {e^{-iq\cdot y} }  \,
\epsilon_a^\lambda(k_2)\,
\epsilon_\mu(q)
\Bigg[
\mathcal G_{\lambda\nu}^{ab}(k_2,x)\,
\bar\psi^i(x)\,
t^b_{ij}\,
\gamma^\nu\,
\mathcal S_{\slashed{\rm P}}^{jk}(x,y)\,
\gamma^\mu\,
\mathcal S^{kl}(y,-k_1)\,
v^l(k_1)
\na
&\hspace{1.2cm}
+
\bar u^l(k_1)\,
\mathcal S^{lk}(k_1,y)\,
\gamma^\mu\,
\mathcal S_{\slashed{\rm P}}^{kj}(y,x)\,
\gamma^\nu\,
t^b_{ji}\,
\psi^i(x)\,
\mathcal G_{\nu\lambda}^{ba}(x,-k_2)
\Bigg].
\end{align}
where \(i,j,k,l,\dots\) are color indices in the fundamental representation, while \(a,b\) are adjoint color indices. The matrices \(t^b\) denote the generators of \(SU(N_c)\) in the fundamental representation, and \(\epsilon_a^\lambda(k_2)\) and \(\epsilon_\mu(q)\) denote the polarization vectors of the outgoing gluon and the incoming virtual photon, respectively.

The second topology is shown in the right panel of Fig.~\ref{fig:qg_dijet_diagram}. In this topology, the target-background insertion is connected directly to the photon vertex, while the final-state gluon is emitted from the outgoing fermion line at a separate vertex. As before, the \(qg\) amplitude contains a charge-conjugate pair, corresponding to an explicit target-background insertion of either \(\bar\psi(x)\) or \(\psi(x)\). The corresponding amplitudes are given by

\begin{align}
\label{eq:fig2_qg_b}
i\mathcal M_{qg}^{(b)}
=&
-ie\,
\lim_{k_1^2,k_2^2\to0}
k_1^2 k_2^2
\int d^{\,4}x\,d^4y\,
e^{-iq\cdot y}\,
\epsilon_a^\lambda(k_2)\,
\epsilon_\mu(q)
\Bigg[
\bar\psi^i(y)\,
\gamma^\mu\,
\mathcal S_{\slashed{\rm P}}^{ij}(y,x)\,
\gamma^\nu t^b_{jk}\,
\mathcal S^{kl}(x,-k_1)\,
v^l(k_1)\,
\mathcal G_{\nu\lambda}^{ab}(k_2,x)
\na
&\hspace{1.2cm}
+
\bar u^l(k_1)\,
\mathcal S^{lk}(k_1,x)\,
\gamma^\nu t^b_{kj}\,
\mathcal S_{\slashed{\rm P}}^{ji}(x,y)\,
\gamma^\mu\,
\psi^i(y)\,
\mathcal G_{\nu\lambda}^{ba}(x,-k_2)
\Bigg].
\end{align}
Equations~\eqref{eq:fig1_qg_a} and \eqref{eq:fig2_qg_b} are the starting point for the quark--gluon channel.  At leading order in the background expansion they generate the bilocal quark operator contribution to the cross section.  Keeping one additional field-strength insertion from any of the propagator blocks generates the quark--gluon--quark trilocal operators needed at twist-three accuracy.

\subsection{Quark-antiquark dijet channel}
\begin{figure}[t]
\begin{center}
\begin{overpic}[width=0.3\textwidth]{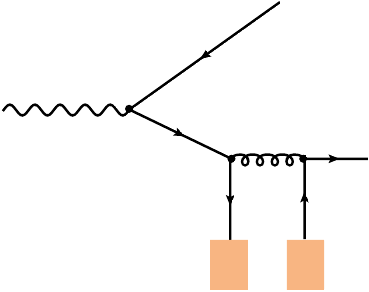}
    \put(3,55){ $ {\gamma^*\,(q)}$} 
    \put(75,80){ $ {k_2}$} 
    \put(100,35){ $ {k_1}$} 
    \put(58, 5){ $\boldsymbol{\bar\psi }$}
     \put(78, 5){ $\boldsymbol{ \psi }$}
\end{overpic}
\end{center}
\caption{Tree-level diagram contributing to the quark--antiquark dijet channel in {the presence of} a quark background. The shaded boxes denote the background antiquark and quark fields, \(\bar{\psi}\) and \(\psi\), respectively. }
\label{fig:qqbar_dijet_diagram}
\end{figure}
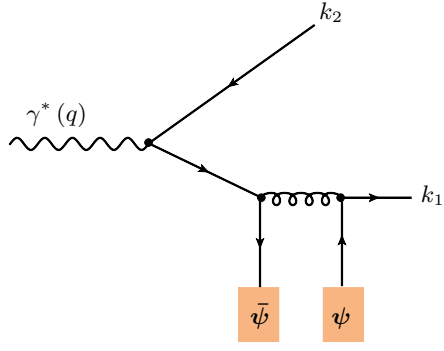

We now turn to the quark--antiquark dijet channel initiated by the quark background.  The tree-level contribution is illustrated in Fig.~\ref{fig:qqbar_dijet_diagram}.  Unlike the quark--gluon channel, this amplitude already contains both background quark fields, \(\psi\) and \(\bar\psi\).  As a result, it contributes to the twist-three cross section through its interference with the leading amplitude containing one gluonic background insertion.  Contributions with additional background quark fields or multiple field-strength insertions are beyond the accuracy considered here and will be omitted.
The corresponding amplitude is given by:
\begin{align}
\label{eq:fig3_qqbar}
i\mathcal M_{q\bar q}
=&-ie\,\epsilon_\rho(q)
\int d^4y\,d^{\,4}z_1\,d^{\,4}z_2\,
e^{-iq\cdot y}
\lim_{k_1^2,k_2^2\to0}
\,
\bar u^i(k_1)
\Big[
\mathcal A_1^\rho+\mathcal A_2^\rho
\Big] 
v^r(k_2),
\end{align}
where
\begin{align}
\mathcal A_1^\rho
=&\,
\mathcal S^{ij}(k_1,z_1)\,
\gamma^\kappa t^a_{jl}\,
\psi^l(z_1)\,
\mathcal G^{ab}_{\kappa\eta}(z_1,z_2)
\bar\psi^m(z_2)\,
t^b_{mn}\,
\gamma^\eta\,
\mathcal S^{nl}_{\slashed{\rm P}}(z_2,y)\,
\gamma^\rho\,
\mathcal S^{lr}(y,-k_2),
\\[0.5em]
\mathcal A_2^\rho
=&\,
\mathcal S^{ij}(k_1,y)\,
\gamma^\rho\,
\mathcal S^{jk}_{\slashed{\rm P}}(y,z_1)\,
\gamma^\kappa t^a_{kl}\,
\psi^l(z_1)
\mathcal G^{ab}_{\kappa\eta}(z_1,z_2)\,
\bar\psi^m(z_2)\,
t^b_{mn}\,
\gamma^\eta\,
\mathcal S^{nr}(z_2,-k_2).
\end{align}

Equations~\eqref{eq:fig1_qg_a}--\eqref{eq:fig3_qqbar} express all tree-level channels in terms of the same small set of background-field propagators.  The next step is therefore common to all the amplitudes above: we evaluate these propagators in the quark background, expand them to the required twist accuracy, and then substitute the resulting building blocks into the amplitudes above.

\section{Background field propagators in quark background}
\label{propagators}
The amplitudes in Sec.~\ref{quark_bg:amp} were written in terms of unevaluated background-field propagators.  We now collect the explicit propagator building blocks that are to be  inserted into those amplitudes.  The derivation follows the method of Ref.~\cite{Mukherjee:2026cte} and is summarized in Appendix~\ref{app:propagators}; here we only  {present} the terms needed for the twist-three cross section.

 {The explicit presence of the quark-background insertions simplify the  {twist}  counting.}  The \(qg\) amplitudes contain one target quark or antiquark field, while the \(q\bar q\) amplitude contains the pair \(\psi\) and \(\bar\psi\).  Therefore, at the  {required} accuracy, the  {remaining} propagators  {may have} at most one additional gluonic field-strength insertion.   {Consequently,} we keep the leading Wilson-line term and the first term proportional to \(F_{-i}\).  The required building blocks are: the LSZ-reduced external quark propagator, the internal quark propagator with its Dirac numerator, and the gluon propagator, which is used both for internal gluon lines and, after {the} LSZ reduction, for the external gluon line.

Throughout this section, fundamental Wilson lines are denoted by \([b^-,a^-]\), while adjoint Wilson lines carry a subscript \(A\), \([b^-,a^-]_A\).  We also define
\begin{align}
\label{eq:sign_function}
s_p \equiv \mathrm{sign}(p^-), 
\qquad
\Theta_p(b^-,a^-)
\equiv
\theta\!\left[s_p(b^- - a^-)\right].
\end{align}

\paragraph{LSZ-reduced quark propagator:}
We first need the fermion propagator attached to an outgoing on-shell quark.  This object appears in the \(qg\) channel and in the \(q\bar q\) channel whenever the external quark leg is LSZ reduced.  For completeness, and to establish our notation, we quote the result of Ref.~\cite{Mukherjee:2026cte} truncated to the single-\(F_{-i}\) precision  {required for} the present calculation:
\begin{equation}
\begin{aligned}
 &\lim_{k_1^2\to 0}  k_1^2\,
  \mathcal S(y,k_1)
\\
&=
-i  e^{-i k_1^+ y^- - i k_1^- y^+}
\: (y_\perp|\bigg\{\!
 i [y^-,\infty]
 + \frac{1}{2k_1^-}\int_{y^-}^{\infty} dx^-\,
 (x^- - y^-)\,
 [y^-, x^-]
 \bigg[2 { k_k} F_{-k} (x^-)
 + \frac{\sigma F(x^-)}{2(x^- - y^-)}
 \bigg]\,
 [x^-,\infty]
 \bigg\}|k_{1\perp})\, .
\end{aligned}
\end{equation}
\\

\paragraph{Internal quark propagator:}
We next need the internal fermion propagator \(\mathcal S_{\slashed{\rm P}}\), which carries the Dirac numerator between interaction vertices.  The derivation, including the separation into the instantaneous and propagating parts with the subsequent ${\rm P}^+$ integration, is given in Appendix~\ref{app:propagators}. Keeping only the terms {at the} required twist accuracy, the gradient-expanded propagator can be written as
\begin{align}
\label{eq:quark_propagator}
 \mathcal S_{\slashed{\rm P}}^{ij}(x,y)\bigg{|}_{x^->y^-}
&=
\int_0^{\infty} {\dhd p^-\over 2p^-}\,
e^{-i p^-(x^+ - y^+)}
 (x_\perp | \:i\, \gamma^- \delta(x^- - y^-) + s_p\,
\Theta_p(x^-,y^-)\,
\mathcal N_q(x^-;p^-,{\rm P_i})
\,
e^{-i (x^- - y^-)
{{\rm P_\perp^2}(x^-)\over 2p^-}}  
\na
&\times \bigg\{\!
-\!i \,[x^-,y^-] 
+
{1\over 2p^-}
\int_{y^-}^{x^-} dx'^-\,
(x'^- - y^-)
[x^-,x'^-]
\left[
2P_k F_{-k} (x'^-)
+
{\sigma F(x'^-)\over 2(x'^- - y^-)}
\right]
[x'^-,y^-]
\bigg\}
 |y_\perp )^{ij},
\end{align}
Here \(\mathcal N_q\) collects the numerator structure of the nonlocal propagating part after the \({\rP^+}\) integration,
\begin{align}
\mathcal N_q(x^-;p^-, {\rm P_i})
\equiv\;&
\gamma^-
\left[
{{\rm P_\perp^2}(x^-)\over 2p^-}
-
{\sigma F(x^-)\over 4p^-}
\right]
+
p^-\gamma^+
-
\gamma_\perp {\rm P_\perp}(x^-).
\label{eq:Nq_definition}
\end{align}
The term proportional to
\(\gamma^- \delta(x^- - y^-)\) is the instantaneous light-front
contribution.  The remaining terms describe propagation through the
background field, with the longitudinal support fixed by
\(s_p\Theta_p(x^-,y^-)\).
\\

\paragraph{Gluon propagator:}
Finally, we quote the gluon propagator in the background field.  In the \(q\bar q\) channel it appears as an internal gluon line, while in the \(qg\) channel the same expression, after LSZ reduction, gives the external gluon leg.  Using the same twist expansion as in Appendix~\ref{app:propagators}, and keeping terms with at most one insertion of the background field strength, we obtain
\begin{align}
\mathcal G^{ab}_{\alpha\beta}(x,y)
&=
-i
\int_{-\infty}^{\infty}
\frac{\dhd p^-}{2p^-}\,
s_p\,
\Theta_p(x^-,y^-)\,
e^{-ip^-(x^+-y^+)}
 (
x_{\perp}
 |
e^{-i(x^- - y^-)
\frac{{\rm P}_\perp^2(x^-)}{2p^-}}
\bigg\{
-i\,g_{\alpha\beta}\,
[x^-,y^-]_A 
\na
&
+
\frac{1}{2p^-}
\int_{y^-}^{x^-} d\xi^-\,
[x^-,\xi^-]_A
\left(
(\xi^- - y^-)
g_{\alpha\beta}
\{P_k,F_{-k}\}(\xi^-)
+
2iF_{\alpha\beta}(\xi^-)
\right)
[\xi^-,y^-]_A
\bigg\}
|
y_{\perp}
)^{ab}.
\label{eq:gluon_prop}
\end{align}
In the actual twist-three amplitude calculation below, the internal gluon propagator is needed only at leading order in the background field, namely through the adjoint Wilson-line term \(-i g_{\alpha\beta}[z_1^-,z_2^-]_A\). We nevertheless {present} the single-field-strength term because the same expression directly generates the LSZ-reduced external gluon line. Assuming \(k_2^- >0\), and following the LSZ procedure described in Appendix F of Ref.~\cite{Mukherjee:2026cte}, we obtain
\begin{align}
\lim_{k_2^2\to0}
k_2^2\,
\mathcal G^{ab}_{\alpha\beta}(k_2,y)
&=\,
i\,e^{ik_2^+y^-+ik_2^-y^+}  (
k_{2\perp}
 |
-i\,g_{\alpha\beta}\,
[\infty,y^-]_A
+
\frac{1}{2k_2^-}
\int_{y^-}^{\infty} d\xi^-\,
[\infty,\xi^-]_A \na
&\times \left[
(\xi^- - y^-)
g_{\alpha\beta}
\,2k_{2k}F_{-k}(\xi^-)
+
2iF_{\alpha\beta}(\xi^-)
\right]
[\xi^-,y^-]_A
 |
y_\perp
 )^{ab}.
\label{eq:amp_gluon_prop_lc}
\end{align}

Substituting these propagators into Eqs.~\eqref{eq:fig1_qg_a}--\eqref{eq:fig3_qqbar} gives the expanded quark-background amplitudes constructed in the next section.

\section{Dijet amplitudes in a quark background}
\label{sec:dijet-amplitudes}
In the previous section, we assembled the background-field building blocks required for the calculation: the internal quark propagator, the LSZ-reduced external quark leg, the internal gluon propagator, and the LSZ-reduced external gluon leg. We now use these blocks to construct the tree-level dijet amplitudes in a quark background.  The operator structures that follow after squaring the amplitudes are discussed in Sec.~\ref{sec:cross_section_operator_structure}.

At twist-three accuracy, the cross section receives contributions from the interference of the leading-twist amplitude with the first twist-suppressed amplitude.  It is therefore sufficient to keep the leading product of background-field blocks and the terms in which one block is replaced by its first subleading correction. 

\subsection{Quark--gluon dijet channel}

The two tree-level topologies in this channel are shown in Fig.~\ref{fig:qg_dijet_diagram}; they differ by whether the final-state gluon is emitted before or after the interaction with the background quark field. {For each topology, } the amplitude is built from two external-line blocks and one topology-dependent internal quark-line block.  The notation used for these blocks is summarized in Table~\ref{tab:qg_subleading_blocks}.

\begin{table}[b]
\centering
\begin{tabular}{c c l}
\hline
Block ~~~~& Topology~~~~ & Origin of the insertion \\
\hline
\({\mathcal X}_0, {\mathcal X}_1\) & \((a),(b)\) & LSZ-reduced external quark line \\
\(\mathcal  {\mathcal H}_0,   {\mathcal H}_1\) & \((a)\) & internal quark line in topology \((a)\) \\
\(\mathcal  {\mathcal L}_0,   {\mathcal L}_1\) & \((b)\) & internal quark line in topology \((b)\) \\
\(\mathcal K_0,\mathcal K_1\) & \((a),(b)\) & LSZ-reduced external gluon line \\
\hline
\end{tabular}
\caption{Leading and subleading insertions retained in the quark--gluon channel at twist-three accuracy.}
\label{tab:qg_subleading_blocks}
\end{table}

The external quark block \({\mathcal X}\) and the external gluon block \(\mathcal K\) appear in both quark--gluon topologies.  The internal block is \(\mathcal H\) for topology \((a)\) and \(\mathcal{L}\) for topology \((b)\), so the two amplitudes contain the products \(\mathcal K \mathcal H {\mathcal X}\) and \(\mathcal K \mathcal  L {\mathcal X}\), respectively.  The subscript \(0\) denotes the leading background-field block, while the subscript \(1\) denotes the same block with exactly one subleading insertion.  The derivation of these blocks, including the required commutator identities, is given in Appendix~\ref{app:qg_building_blocks}.  Here we use the resulting expressions to build the two amplitudes.

For each topology, we {only show} the contribution proportional to the background field \(\bar\psi\); the corresponding \(\psi\) contribution is obtained by charge conjugation.

\subsubsection{Topology (a)} 

We first consider the topology (a) contribution in Fig.~\ref{fig:qg_dijet_diagram} with the explicit background-field insertion $\bar\psi(x)$.  Starting from
Eq.~\eqref{eq:fig1_qg_a}, the amplitude can be organized in terms of three amputated building blocks: the LSZ-reduced gluon kernel $\mathcal K_{\lambda\nu}^{ac}$, the dressed internal quark-line block $\mathcal  H^{\nu c}$, and the amputated external quark block ${\mathcal X}$.  After performing the $x^+$ and $p^-$ integrations, we obtain
\begin{align}
\label{eq:qg_a_barpsi_integrated}
i\mathcal M_{qg,\bar\psi}^{(a)}
&=
e
\int dy^-\, d^2 y_\perp
\int_{y^-}^{\infty} dx^-
\frac{
e^{i k_q^+ y^- - i k_{q\perp}\cdot y_\perp}
}
{(2k_2^-)(2q^-)}
\epsilon_a^\lambda(k_2)\,
\epsilon_\mu(q)\,
\mathcal K_{\lambda\nu}^{ac}(k_2;x^-,y_\perp)\,
\mathcal  H^{\nu c}(x^-,y^-;y_\perp;k_2)
\gamma^\mu
{\mathcal X}(k_1,y)
v(k_1),
\end{align}
where we set \(p=k_2\) in the dressed quark-line block. 
The integral over $y^+$ results in the factor  \(
2\pi\delta(q^- - k_1^- - k_2^-)
\). 
To maintain consistency with the cross section normalization adopted throughout this work, we replace this factor with  \(1/(2q^-)\) in Eq.~\eqref{eq:qg_a_barpsi_integrated}. Here
\(y_\perp\) denotes the position of the Wilson lines and background field insertions within any given building block.

Each building block in Eq.~\eqref{eq:qg_a_barpsi_integrated}  is twist-expanded to an appropriate order 
\begin{align}
\label{eq:decomposition}
\mathcal K_{\lambda\nu}^{ac}
&=
\mathcal K_{0,\lambda\nu}^{ac}
+
\mathcal K_{1,\lambda\nu}^{ac}
+\cdots ,
\\
\mathcal  H^{\nu c}
&=
\mathcal  {\mathcal H}_0^{\nu c}
+
\mathcal  {\mathcal H}_1^{\nu c}
+\cdots ,
\\
{\mathcal X}
&=
{\mathcal X}_0+{\mathcal X}_1+\cdots .
\end{align}
The subscript $0$ denotes the leading twist terms, while the subscript $1$ contains terms which  are sub-leading in dynamical twist (contain a $F_{-i}$ insertion) or are sub-leading in kinematic twist (at $\mathcal{O}(\Delta/P)$ order). Thus, restricting ourselves to twist-three accuracy in the cross-section, the product of building blocks in Eq.~\eqref{eq:qg_a_barpsi_integrated} can be expanded as
\begin{align}
\mathcal K \mathcal  H {\mathcal X}
\big|_{\mathrm{tw}\leq 3}
=
\mathcal K_0 \mathcal  {\mathcal H}_0 {\mathcal X}_0
+
\mathcal K_1 \mathcal  {\mathcal H}_0 {\mathcal X}_0
+
\mathcal K_0   {\mathcal H}_1 {\mathcal X}_0
+
\mathcal K_0   {\mathcal H}_0 {\mathcal X}_1 .
\end{align}
Terms with two or more subleading insertions, such as
$\mathcal K_1 \mathcal  {\mathcal H}_1$, $\mathcal K_1{\mathcal X}_1$, or $\mathcal  {\mathcal H}_1{\mathcal X}_1$, are beyond the
twist-three accuracy.

For the quark and gluon operator, it is convenient to introduce Wilson-line-dressed background fields
\begin{align}
\bar\Psi(x^-)
&\equiv
\bar\psi(x^-)\,[x^-,\infty],
\\
\Psi(x^-)
&\equiv
[\infty,x^-]\,\psi(x^-),
\\
\bar F_{\alpha\beta}(x^-)
&\equiv
[\infty,x^-]\,
F_{\alpha\beta}(x^-)\,
[x^-,\infty],
\end{align}
where the transverse coordinate $y_\perp$ is suppressed.  We also use the short-hand notation
\begin{equation}
\sigma\bar F
\equiv
\sigma^{\alpha\beta}\bar F_{\alpha\beta}.
\end{equation}
We now explicitly write down the building blocks defined in Eq.(\ref{eq:decomposition}). Starting with $K_{\lambda\nu}^{ac}$, the LSZ-reduced gluon kernel is
\begin{equation}
\mathcal K_{\lambda\nu}^{ac}
=
\mathcal K_{0,\lambda\nu}^{ac}
+
\mathcal K_{1,\lambda\nu}^{ac}
+\cdots ,
\end{equation}
with
\begin{align}
\mathcal K_{0,\lambda\nu}^{ac}(k_2;x^-)
&=
g_{\lambda\nu}\,\delta^{ac},
\\
\mathcal K_{1,\lambda\nu}^{ac}(k_2;x^-)
&=
\frac{i}{2k_2^-}
\int_{x^-}^{\infty} dz_1^-\,
(z_1^- - x^-)
\left[
[\infty,z_1^-]_{A}
\left(
2 g_{\lambda\nu}\, k_{2r} F_{-r}(z_1)
+
\frac{2i F_{\lambda\nu}(z_1)}
{z_1^- - x^-}
\right)
[z_1^-,\infty]_{A}
\right]^{ac}.
\label{eq:K1_gluon_insertion}
\end{align}
Here the Wilson lines in $\mathcal K$ are in the adjoint representation.

The dressed quark-line block is
\begin{align} 
\mathcal  H^{\nu c}
&=
\mathcal  {\mathcal H}_0^{\nu c}
+
\mathcal  {\mathcal H}_1^{\nu c}
+\cdots ,
\end{align}
with
\begin{align}
\label{eq:H0_def_main}
  {\mathcal H}_0^{\nu c}(x^-,y^-;p)
&=
-i\,
\bar\Psi(x^-)\,
\gamma^\nu
t^c
\,T_0(x^-,y^-;p),\na
  {\mathcal H}_1^{\nu c}(x^-,y^-;p)
&=
\left[
i\,D_i\,\bar\psi(x)\,[x^-,\infty]
\right]
\gamma^\nu\,
T_i\,
t^c
+
\bar\Psi(x^-)\,
\gamma^\nu\,
T_i \, \left(
\int_{x^-}^{\infty} d\xi^-\,
\bar F_{-i}(\xi^-)\,
t^c + t^c \int_{x^-}^{\infty} d\xi^-\,
\bar F_{-i}(\xi^-) \right)\na
&+
\frac{i}{4p^-}
\bar\Psi(x^-)
\gamma^\nu
t^c
\gamma^-\sigma\bar F(x^-)
+
\bar\Psi(x^-)\,
\gamma^\nu
t^c
T_0\,
\frac{1}{2p^-}
\int_{y^-}^{x^-} d\xi^-\,
(\xi^- - y^-)
\left[
2p_k\,\bar F_{-k}(\xi^-)
+
\frac{\sigma\bar F(\xi^-)}
{2(\xi^- - y^-)}
\right],
\end{align}
where
\begin{align}
\label{eq:Gamma0_def_main}
T_0(x^-,y^-;p)
= &
\left(
i\delta(x^- - y^-)
+
\frac{p_\perp^2}{2p^-}
\right)\gamma^-
+
p^-\gamma^+
+
p_\perp^r\gamma_r ,\na 
T_i(x^-,y^-;p)
= &
\frac{(x^- - y^-)\,p_i}{p^-}\,
T_0(x^-,y^-;p)
+
i\left(
\frac{p_i}{p^-}\gamma^-
+
\gamma_i
\right).
\end{align}
 The first three terms in \(\mathcal{H}_1^{\nu c}\) are generated by commuting the covariant momenta $P_i$ in the internal quark block $\mathcal S_{\slashed{\rm P}}^{ij}(x,y)$, across the LSZ-reduced gluon propagator $\mathcal G_{\nu\lambda}^{ab}(k_2,x)$, while the fourth and the fifth terms come directly from the sub-leading pieces of $\mathcal S_{\slashed{\rm P}}^{ij}(x,y)$\par

Finally, the amputated external quark block is given by 
\begin{align}
{\mathcal X}(k_1,y)
=
{\mathcal X}_0(k_1,y)
+
{\mathcal X}_1(k_1,y)
+\cdots ,
\end{align}
with
\begin{align}
\label{eq:A0_def_main}
{\mathcal X}_0(k_1,y)
&=
1,
\\
\label{eq:A1_def_main}
{\mathcal X}_1(k_1,y)
&=
-\frac{i}{2k_1^-}
\int_{y^-}^{\infty} dz_1^-\,
(z_1^- - y^-)
\left[
2k_{1r}\,\bar F_{-r}(z_1^-)
-
\frac{\sigma\bar F(z_1^-)}
{2(z_1^- - y^-)}
\right].
\end{align}

Equations~\eqref{eq:qg_a_barpsi_integrated}--\eqref{eq:A1_def_main}
are the topology-\(a\) ingredients from the background-field expansion.

The contribution with an explicit background $\psi(x)$ insertion is obtained by charge conjugation.  It has the same building-block structure as Eq.~\eqref{eq:qg_a_barpsi_integrated}, with the ordering of the fermion-line matrices and color generators reversed according to the charge-conjugation rules summarized in Appendix~\ref{app:charge_conjugation_rules}.

\subsubsection{Topology (b)}

We next consider topology~(b).  Starting from Eq.~\eqref{eq:fig2_qg_b}, and using the same universal external-line building blocks \(\mathcal K\) and \({\mathcal X}\) introduced in topology~(a), the contribution proportional to the background field \(\bar\psi(y)\) can be written as
\begin{align}
\label{eq:qg_b_barpsi}
i\mathcal M_{qg,\bar\psi}^{(b)}
=&
-ie
\int dy^-\, d^2y_\perp
\int_{y^-}^{\infty} dx^-\,
\frac{
e^{ik_q^+ y^- - i k_{q\perp}\cdot y_\perp}
}
{4(k_1^-+k_2^-)q^-}
\,
\exp\left[
i
\frac{(k_1+k_2)^2}
{2(k_1^-+k_2^-)}
(x^- - y^-)
\right]\epsilon_a^\lambda(k_2)\,
\epsilon_\mu(q)\,
\nonumber\\[4pt]
&\times
\bar\Psi^o(y)\,
\gamma^\mu\,
\mathcal L^{om}(x^-,y^-;y_\perp;-(k_1+k_2))
\gamma^\nu\,
\mathcal X^{nl}(k_1,x^-,y_\perp)\,
v^l(k_1) 
\mathcal K_{\nu\lambda}^{ac}(k_2;x^-,y_\perp)\,
t^c_{mn}.
\end{align}
Here the internal momentum flow through the topology-(b) quark-line block
is  $p=-(k_1+k_2)$.
As in topology~(a), the \(y^+\) integration was performed and the
corresponding longitudinal delta function was removed according to
our cross section normalization convention, leaving the explicit factor
\(1/(2q^-)\) in Eq.~\eqref{eq:qg_b_barpsi}.  The transverse
coordinate \(y_\perp\) denotes the common transverse position of the Wilson
lines and background fields.

The only new building block in topology~(b) is the internal quark-line block
\(\mathcal L\).  It is expanded as
\begin{align}
\mathcal L(x^-,y^-;p)
=
  {\mathcal L}_0(x^-,y^-;p)
+
  {\mathcal L}_1(x^-,y^-;p)
+\cdots ,
\end{align}
where the leading block is
\begin{align}
\label{eq:L0_def_main}
  {\mathcal L}_0(x^-,y^-;p)
=
T_0^*(x^-,y^-;p)=\left(
-i\delta(x^- - y^-)
+
\frac{p_\perp^2}{2p^-}
\right)\gamma^-
+
p^-\gamma^+
+
p_\perp^r\gamma_r ,
\end{align}
and the subleading-insertion block is
\begin{align}
\label{eq:L1_def_main}
  {\mathcal L}_1(x^-,y^-;p)
&=
-i\,
\int_{x^-}^{\infty} d\xi^-\,
\bar F_{-i}(\xi^-) \,
\left(\frac{(x^- - y^-)\,p_i}{p^-}\,
\slashed{p}
-
i\left(
\frac{p_i}{p^-}\gamma^-
+
\gamma_i
\right) \right)
-
\frac{\sigma\bar F(x^-)\gamma^-}{4p^-}\na
&-
i\,\frac{1}{2p^-}
\int_{y^-}^{x^-} d\xi^-\,
(\xi^- - y^-)
\left[
2p_k\,\bar F_{-k}(\xi^-)
+
\frac{\sigma\bar F(\xi^-)}
{2(\xi^- - y^-)}
\right]\,
T_0(x^-,y^-;p)\,.
\end{align}
To twist-three accuracy, the product of building blocks in Eq.~\eqref{eq:qg_b_barpsi} reads 
\begin{align}
\mathcal K \mathcal L {\mathcal X}
\big|_{\mathrm{tw}\leq 3}
=
\mathcal K_0   {\mathcal L}_0 {\mathcal X}_0
+
\mathcal K_1   {\mathcal L}_0 {\mathcal X}_0
+
\mathcal K_0   {\mathcal L}_1 {\mathcal X}_0
+
\mathcal K_0   {\mathcal L}_0 {\mathcal X}_1 .
\end{align}

The contribution with an explicit background \(\psi(y)\) insertion is obtained by charge conjugation, see  Appendix~\ref{app:charge_conjugation_rules}.  We denote the corresponding amplitude by \(i\mathcal M_{qg,\psi}^{(b)}\).

\subsection{Quark--antiquark dijet channel}
\label{subsec:qqbar-amplitude}

We now consider quark--antiquark dijet production in the background quark field.  This channel is organized slightly differently from the quark--gluon channel.   The required amplitude ingredients are: the contribution with an explicit background \(\psi\bar\psi\) pair and the contribution with one background field-strength insertion.    The relevant diagrams are shown in Fig.~\ref{fig:qqbar_dijet_diagram}.

We first evaluate the contribution with the background-field insertion on the quark line,
\begin{align}
\label{eq:fig3_qqbar_1}
i\mathcal M_{q\bar q}^{(q)}
&=-ie\,\epsilon_\rho(q)
\int d^4y\,d^{\,4}z_1\,d^{\,4}z_2\,
e^{-iq\cdot y}\na
&\times
\lim_{k_1^2,k_2^2\to0}
\,
\bar u^i(k_1)
\Big[
\mathcal S^{ij}(k_1,z_1)\,
\gamma^\kappa t^a_{jl}\,
\psi^l(z_1)\,
\mathcal G^{ab}_{\kappa\eta}(z_1,z_2)
\bar\psi^m(z_2)\,
t^b_{mn}\,
\gamma^\eta\,
\mathcal S^{no}_{\slashed{\rm {P}}}(z_2,y)\,
\gamma^\rho\,
\mathcal S^{or}(y,-k_2)
\Big] 
v^r(k_2),
\end{align}
where  the background quark fields are explicitly indicated by \(\psi(z_1)\) and \(\bar\psi(z_2)\).  Since this already gives the two-quark background insertion required for the cross section, the remaining background-field propagators only need to be kept at leading order in the background gauge field.  Substituting the leading twist expressions for the LSZ-reduced external quark line, the internal gluon propagator, and the internal quark propagator, the plus-coordinate integrations can be performed straightforwardly. 
We obtain
\begin{align}
 {i\mathcal M_{q\bar q}^{(q)}} 
&=
-\frac{e\,\epsilon_\rho(q)}{2 q^- \,k_q^+}
\int d^2y_\perp
\int  dz_1^-
\int_{-\infty}^{z_1^-} dz_2^-\,
e^{ik_q^+ z_2^- - ik_{q\perp}\cdot y_\perp}
\bar u^i(k_1)\,
\gamma^\kappa
\psi^\ell(z_1^-,y_\perp)\,
\mathcal T^{io}_{\ell m}(z_1^-,z_2^-;y_\perp)\,
\bar\psi^m(z_2^-,y_\perp)
\na
&\times
\frac{1}{2k_1^-}
\gamma_\kappa
\left[
\gamma^-
\left(
ik_q^+
+
\frac{k_{1\perp}^2}{2k_1^-}
\right)
+
\gamma^+ k_1^-
-
\gamma_i k_{1i}
\right]
\gamma^\rho
v^o(k_2),
\label{eq:qqbar_A1_before_color}
\end{align}
where the color structure is
\begin{align}
\mathcal T^{io}_{\ell m}(z_1^-,z_2^-;y_\perp)
=
[\infty,z_1^-]^{ij}_{y_\perp}
t^a_{j\ell}
[z_1^-,z_2^-]^{ab}_{A,y_\perp}
t^b_{mn}
[z_2^-,\infty]^{no}_{y_\perp}.
\label{eq:qqbar_color_kernel_def}
\end{align}
The adjoint Wilson line in Eq.~\eqref{eq:qqbar_color_kernel_def} can be transformed into fundamental Wilson lines via the Fierz identity.  Using
\begin{align}
t^a_{j\ell}
[z_1^-,z_2^-]^{ab}_{A}
t^b_{mn}
=
\frac{1}{2}
\left(
[z_1^-,z_2^-]^{jn}
[z_2^-,z_1^-]^{m\ell}
-
\frac{1}{N_c}
\delta_{j\ell}\delta_{mn}
\right),
\label{eq:qqbar_fierz_identity}
\end{align}
we get
\begin{align}
  \mathcal T^{io}_{\ell m}(z_1^-,z_2^-;y_\perp)
= \delta^{io}
[z_2^-,z_1^-]^{m\ell}_{y_\perp}
-
\frac{1}{N_c}
[\infty,z_1^-]^{i\ell}_{y_\perp}
[z_2^-,\infty]^{mo}_{y_\perp}.
\end{align}

The contribution with the background-quark insertion attached to the outgoing antiquark line is obtained analogously.  Equivalently, it follows from Eq.~\eqref{eq:qqbar_A1_before_color} by applying charge conjugation to the fermion line and reversing the ordering of the corresponding Wilson-line operators.  We denote this contribution by \(i\mathcal M_{q\bar q}^{(\bar q)}\).  The full two-quark-background amplitude is then
\begin{align}
i\mathcal M_{q\bar q}^{(\psi\bar\psi)}
=
i\mathcal M_{q\bar q}^{(q)}
+
i\mathcal M_{q\bar q}^{(\bar q)} .
\label{eq:qqbar_psipsi_amplitude}
\end{align}
This result is used below to construct the interference with the corresponding leading-power amplitude and to identify the associated two-quark background operator structure.

The second ingredient needed to compute the cross-section in this channel is the amplitude with a single background field-strength insertion derived in our previous work~\cite{Mukherjee:2026cte}.  In the notation used here, it  reads
\begin{align}
 i\mathcal{M}_{q \bar q}^{(F)}
 =&- e\,\epsilon_\rho(q)
\int \frac{dx^-  d^2 x_\perp}{2q^-}
e^{i\left(k_q^+ x^-- k_{q\perp} \cdot  {x}_\perp\right)}  \,
\bar{u}(k_1)
\bigg\{
  \bb{C}_1^i \gamma^\rho
  \overline{F}_{-i}(x^-,x_\perp)
+
 i\left(
 \bb{S}_1 \sigma^{\mu\nu}\gamma^\rho
 + \bb{S}_2 \gamma^\rho\sigma^{\mu\nu}
 \right)
 \overline{F}_{\mu\nu}(x^-,x_\perp)
\bigg\}
v(k_2),
\label{eq:Mqqbar_F}
\end{align}
where the coefficients \( \bb{C}_1^i\),\,\(\bb{S}_1 \), and \(\bb{S}_2\) can be found in Ref~\cite{Mukherjee:2026cte}.

We thus collected all amplitude-level ingredients needed at twist-three accuracy.  The quark--gluon channel gives the two topologies \((a)\) and \((b)\), together with their \(\psi\) and \(\bar\psi\) background insertions and the single subleading insertions in \({\mathcal X}\), \(\mathcal K\), \(\mathcal  H\), or \(\mathcal L\).  The quark--antiquark channel  gives the two-quark-background amplitude in Eq.~\eqref{eq:qqbar_psipsi_amplitude} and the field-strength amplitude in Eq.~\eqref{eq:Mqqbar_F}.  We now compute the cross section by squaring these amplitudes and grouping the resulting terms by the background operators.

\section{Cross section and operator structure}
\label{sec:cross_section_operator_structure}

The cross section is organized according to the operator structures.  We first consider the bilocal sector, which contains two background quark fields.  We then turn to the trilocal sector, where the two background quark fields are accompanied by one field-strength insertion.

For each photon polarization \(\lambda=L,T\), the quantity entering
Eq.~\eqref{eq:xsec_channel_general} can be written as
\begin{align}
\mathcal W^\lambda_{\rm tw\leq 3}
=
\left|i\mathcal M_{qg}^{\lambda,(2)}\right|^2
+
2\,{\rm Re}\,\left[
i\mathcal M_{qg}^{\lambda,(3)}
\left(i\mathcal M_{qg}^{\lambda,(2)}\right)^\dagger
\right]
+
2\,{\rm Re}\,\left[
i\mathcal M_{q\bar q}^{\lambda,(F)}
\left(i\mathcal M_{q\bar q}^{\lambda,(\psi\bar\psi)}\right)^\dagger
\right].
\label{eq:cross_section_sector_decomposition}
\end{align}
Here, the labels \((2)\) and \((3)\) refer to the corresponding contributions at the amplitude level and should not be confused with the bilocal and trilocal operator sectors of the cross section.  The first squared amplitude contains two background quark fields and therefore generates bilocal quark operators, schematically of the form
\(
\bar\psi\,\Gamma\,\psi ,
\)
where \(\Gamma\) denotes the Dirac structure appropriate to the operator under consideration.
By contrast, the interference between \(i\mathcal M_{qg}^{\lambda,(3)}\) and
\(i\mathcal M_{qg}^{\lambda,(2)}\) contains two background quark fields together with one field-strength insertion.  It therefore generates trilocal quark--gluon--quark operators, schematically of the form
\(
\bar\psi\,\Gamma\,F_{-i}\,\psi .
\)
The last term is the analogous interference contribution in the \(q\bar q\) channel, between the amplitude containing two background quark fields and the amplitude containing a single field-strength insertion.  It therefore also belongs to the trilocal sector.  The overall phase-space normalization is fixed by Eq.~\eqref{eq:xsec_channel_general}.

We present the results separately for longitudinally and transversely polarized virtual photons.  For longitudinal polarization, we use the convention
\begin{align}
\epsilon_\mu^L(q)
    =
    \frac{Q}{q^-}\, g_{\mu +} \,. 
\end{align}
For transverse polarization, we average over the two physical transverse states and use
\begin{align}
\frac{1}{2}
\sum_{\lambda=\pm 1}
\epsilon_k^\lambda(q)\,
\epsilon_l^{\lambda *}(q)
=
-\frac{1}{2}\,g_{kl},
\qquad
k,l=1,2 .
\end{align}

\subsection{ {Bilocal operators}} 

We now turn to the bilocal contribution to the quark--gluon dijet cross section.  This part is obtained by retaining only the leading terms in the background-field expansion.  For the first topology, this corresponds to keeping the \({\mathcal X}_0\), \(\mathcal{K}_0\), and \(  {\mathcal H}_0\) pieces, while for the second topology we keep the \({\mathcal X}_0\), \(\mathcal{K}_0\), and \(\mathcal  {\mathcal L}_0\) pieces. Since none of these terms contains an explicit field-strength insertion, their interference with the corresponding charge-conjugate amplitudes produces the
bilocal quark operator
\begin{align}
\mathcal O^{\mu}
&\equiv
\bar\Psi(y^-,y_\perp) \,
\gamma^\mu\,
\Psi(x^-,x_\perp)
+
\bar\Psi(x^-,x_\perp) \,
\gamma^\mu\,
\Psi(y^-,y_\perp) .
\label{eq:twobody_operator}
\end{align}
The projection onto \(\mathcal O^+\) gives the leading-twist contribution, whereas the projection onto \(\mathcal O^{\,i}\) gives the dynamical twist-three bilocal contribution.

For the first topology, the leading contribution with a background quark field is

\begin{align}
\label{eq:qg_barpsi_a_2_final}
i\mathcal M_{qg,\bar\psi}^{(a,2)}
&=
-e
\int \frac{dx^-  d^2 x_\perp}{2q^-}
e^{i\left(k_q^+ x^-- k_{q\perp} \cdot  {x}_\perp\right)} 
\epsilon_a^\lambda(k_2) \,
\epsilon_\mu(q)\,
\bar\Psi^n(x^-,x_\perp) \,
t^a_{nl} \,
\gamma_\lambda \gamma_\nu \, \mathcal{B}^{\nu} (k_1,k_2,q)
\gamma^\mu
v^l(k_1),
\end{align}
with
\begin{align}
\mathcal B^+ 
&=
\frac{k_2^+ - k_q^+}{2k_2^- k_q^+}\approx
-
\frac{
\bar z  P_\perp^2+\epsilon_f^2
}{
2\bar z\,q^- {{\tilde P^2}} 
}
-
\frac{
z\,(\Delta \!\cdot\! P) 
}{
q^- {{\tilde P^2}} 
}
+O(\Delta_\perp^2),
\\[6pt]
\mathcal B^- 
&=
\frac{1}{2k_q^+}
\approx
\frac{
z\bar z\,q^-
}{
 {{\tilde P^2}} 
}
+O(\Delta_\perp^2),
\\[6pt]
\mathcal B^{\,i} 
&=
\frac{k_2^i}{2k_2^- k_q^+}
 \approx
-
\frac{
z\,P_\perp^i
}{
 {{\tilde P^2}} 
}
+
\frac{
z\bar z\,\Delta_\perp^i
}{
 {{\tilde P^2}} 
}
+O(\Delta_\perp^2).
\end{align}
where we  introduced the shorthand notation 
\begin{equation}
    \tilde P^2 \equiv P_\perp^2+\epsilon_f^2
\end{equation}  {with \(\epsilon_f^2 = z\bar z Q^2\)},  $z = k_1^-/q^-$ and $\bar z = k_2^-/q^- = 1-z$. The contribution with a background antiquark field is obtained by charge conjugation. Here and in what follows, the approximate equalities are obtained by expanding the back-to-back kinematics in the small momentum imbalance \(\Delta_\perp/P_\perp\).  We keep terms through linear order in \(\Delta_\perp\) and neglect corrections of \(O(\Delta_\perp^2)\).

For the second topology, keeping only the ${\mathcal X}_0$, $\mathcal{K}_0$ and $  {\mathcal L}_0$ building blocks gives
\begin{align}
\label{eq:qg_barpsi_b_2_final}
i\mathcal M_{qg,\bar\psi}^{(b,2)}
&=
- e
\int \frac{dx^-  d^2 x_\perp}{2q^-}
e^{i\left(k_q^+ x^-- k_{q\perp} \cdot  {x}_\perp\right)} 
\epsilon_\mu(q)\,
\epsilon_a^\lambda(k_2)\,
\bar\Psi^o(x^-,x_\perp)\,
\gamma^\mu
\gamma_\nu \mathcal{D}^\nu (k_1,k_2)
\gamma_\lambda
v^l(k_1)\,
t^a_{ol}.
\end{align}
with
\begin{align}
\mathcal D^+ 
&=
\frac{
k_1^+ + k_2^+
}
{(k_1+k_2)^2}
\approx
\frac{1}{2q^-}
+O(\Delta_\perp^2),
\\[6pt]
\mathcal D^- 
&=
\frac{
k_1^- + k_2^-
}
{(k_1+k_2)^2}
\approx
\frac{
z\bar z\,q^-
}{
P_\perp^2
}
+O(\Delta_\perp^2),
\\[6pt]
\mathcal D^{\,i}  
&=
\frac{
k_1^i + k_2^i
}
{(k_1+k_2)^2}
\approx
\frac{
z\bar z\,\Delta_\perp^i
}{
P_\perp^2
}
+O(\Delta_\perp^2).
\end{align}
The charge-conjugate contribution is obtained by reversing the fermion-line ordering.

The complete bilocal amplitude in the quark--gluon channel is therefore the sum of the two topologies and their charge-conjugate counterparts,
\begin{align}
\label{eq:qg_2_total_amplitude}
i\mathcal M_{qg}^{(2)}
&=
i\mathcal M_{qg,\bar\psi}^{(a,2)}
+
i\mathcal M_{qg,\psi}^{(a,2)}
+
i\mathcal M_{qg,\bar\psi}^{(b,2)}
+
i\mathcal M_{qg,\psi}^{(b,2)}\na [4pt]
&
  = e \int \frac{dx^-  d^2 x_\perp}{2q^-}
e^{i\left(k_q^+ x^-- k_{q\perp} \cdot  {x}_\perp\right)} \left( \epsilon_i^{\lambda}(k)\,\epsilon_\mu(q) \, \mathcal{B}^\nu  + \epsilon_{i\mu}(k)\,\epsilon^\lambda(q)\, \mathcal{D}^\nu \right) \, t^i_{de} \bar{u}^d(k_1) \gamma^\mu\,\gamma_\nu\,\gamma_{\lambda}\, [\infty,x^-]^{ec}_{x_\perp}\psi^c(x^-,x_\perp)\na [4pt]& 
{\, -}\,e \int \frac{dx^-  d^2 x_\perp}{2q^-}
e^{i\left(k_q^+ x^-- k_{q\perp} \cdot  {x}_\perp\right)} \left( \epsilon_i^{\lambda}(k)\:\epsilon_\mu(q)\, \mathcal{B}^\nu  + \epsilon_{i\mu}(k)\:\epsilon^\lambda(q) \, \mathcal{D}^\nu \right)  \:t^i_{ed}\:\bar{\psi}^c(x^-,x_\perp) [x^-,\infty]^{ce}_{x_\perp}\:\:\gamma_{\lambda}\: 
\gamma_\nu\:\gamma^\mu\:\:v^d(k_1) 
\end{align}
The bilocal contribution to the cross section is then obtained by multiplying this amplitude with its complex conjugate,
\begin{equation}
\label{eq:qg_twobody_cross_section}
\left.
d\sigma_{qg}
\right|_{\rm biloc.}
\propto
\sum_{\text{pol., color}}
\mathcal M_{qg}^{(2)}
\left[
\mathcal M_{qg}^{(2)}
\right]^\dagger .
\end{equation} 
After carrying out the Dirac algebra, we project the resulting quark operator onto the vector channel and express the result in terms of
\(\mathcal O_\mu\) defined in Eq.~\eqref{eq:twobody_operator}.  In addition, the kinematical twist-three contribution is obtained by expanding the leading-twist hard coefficient to first order in
\(\Delta_\perp/P_\perp\).

\subsubsection{ {Longitudinally polarized photon}}

We first consider the longitudinally polarized virtual photon.  Using the polarization convention introduced above, the two photon vertices are projected onto the minus direction.  After projecting the resulting bilocal interference onto the vector channel,
\begin{align}
\Gamma\Big|_{\rm vec}
=
\frac14
{\rm Tr}\!\left(\gamma_\rho \Gamma\right)
\gamma^\rho ,
\end{align}
the \(\rho=+\) component gives the leading-twist operator \(\mathcal O^+\), whereas the transverse component \(\rho=i\) gives the dynamical twist-three operator \(\mathcal O^{\,i}\).  The kinematical twist-three contribution from the \(\Delta_\perp/P_\perp\) expansion of the leading-twist coefficient vanishes for longitudinal polarization.

Combining these contributions, the longitudinal bilocal result through twist-three accuracy is
\begin{align}
\left.|i\mathcal M^L_{qg}|^2\right|_{\rm biloc.}
&=\,
2\, e^2\, C_F
\int
dx^- d^2x_\perp\,
dy^- d^2y_\perp\,
e^{i k_q^+ (x^- - y^-)
-i \Delta_{\perp}\cdot( x_\perp - y_\perp)}\na
&\times \left[
\frac{\bar z z^2 \epsilon_f^2}
{q^-  {{\tilde P^4}} }
\mathcal O^+
+
z^2\epsilon_f^2
\frac{
\left(P_\perp^2+{{\tilde P^2}}\right) P_\perp^i
}{
(q^-)^2 P_\perp^2  {{\tilde P^4}}
}
\mathcal O^{\,i}
\right]
+
\left.|i\mathcal M^L_{qg}|^2\right|_{\rm kin.\,3}^{(3\to2)}.
\label{eq:qg_L_twobody_total}
\end{align}

The leading twist part of this result can be compared to its small-$x$ counterpart available in literature. 
To take the small-$x$ limit we consider the kinematic region where  the longitudinal momentum fraction carried by the target field entering the hard scattering is small,
\[
x \equiv \frac{k_q^+}{P^+}\ll 1 ,
\]
or equivalently \(k_q^+\to 0\) at fixed target momentum \(P^+\). In this limit, the first term in the square bracket of Eq.~\eqref{eq:qg_L_twobody_total} reduces to  Eq.~(46) of Ref.~\cite{Altinoluk:2023qfr}. Note that calculations in Ref.~\cite{Altinoluk:2023qfr} were performed to leading-twist precision only.  

The remaining terms in Eq.~\eqref{eq:qg_L_twobody_total} are twist-three contributions. They have two distinct origins. The second term in the square  bracket of Eq.~\eqref{eq:qg_L_twobody_total} is a dynamical twist-three contribution associated with the transverse-vector projection, whereas the last term in Eq.~\eqref{eq:qg_L_twobody_total} is a kinematical twist-three contribution arising from the contact part of the trilocal expansion.  
This contact contribution reduces to an effective bilocal term, as will
be shown explicitly in Eq.~\eqref{eq:qg_longitudinal_ct}. We denote it by the superscript \((3\to2)\) to emphasize its trilocal origin.

\subsubsection{ {Transversely polarized photon} }

  As in the
longitudinal case, we project the bilocal interference onto the vector channel.  The \(\rho=+\) component gives the leading-twist operator \(\mathcal O^+\), while the transverse component \(\rho=i\) gives the dynamical twist-three operator \(\mathcal O^{\,i}\).  The kinematical twist-three term is obtained by expanding the \(\gamma^+\)-projected hard coefficient to first order in \(\Delta_\perp/P_\perp\).

Combining the leading-twist contribution, the kinematical twist-three
correction, and the dynamical twist-three contribution from the transverse vector projection, we obtain
\begin{align}
\left.|i\mathcal M^T_{qg}|^2\right|_{\rm biloc.}
=&\,
e^2 \,C_F
\int
dx^- d^2x_\perp\,
dy^- d^2y_\perp\,
e^{i k_q^+ (x^- - y^-)
-i \Delta_{\perp}\cdot( x_\perp - y_\perp)}
\Bigg\{
\frac{z}{2q^-}
\left[
\frac{z^2+1}{P_\perp^2}
-
\frac{\epsilon_f^2(z^2+\bar z^2)}
{ {{\tilde P^4}} }
-
\frac{z-\bar{z}}{ {{\tilde P^2}} }
\right]
\mathcal O^+
\na
& 
+
\frac{z^2\bar z}{q^-}
\frac{
(\Delta \cdot P )
\left(\bar z P_\perp^2+z\epsilon_f^2\right)
}{
P_\perp^2
 {{\tilde P^4}} 
}
\mathcal O^+ 
-
\frac{z\,P_\perp^i}{2(q^-)^2}
\left[
\frac{
P_\perp^2(z-\bar z)+\dfrac{z}{\bar z}\epsilon_f^2
}
{
 {{\tilde P^4}} 
}
+
\frac{
\epsilon_f^2
}
{
P_\perp^2 {{\tilde P^2}} 
}
\right]
\mathcal O^{\,i}
\Bigg\}
+
\left.|i\mathcal M^T_{qg}|^2\right|_{\rm kin.\,3}^{(3\to2)}
.
\label{eq:qg_T_twobody_total}
\end{align}
The three terms in braces correspond respectively to the leading-twist contribution, the kinematical twist-three correction from the \(\Delta_\perp/P_\perp\) expansion of the leading-twist coefficient, and the dynamical twist-three contribution from the transverse vector projection. In the small-\(x\) limit, the leading-twist term reproduces Eq.~(47) of Ref.~\cite{Altinoluk:2023qfr}. The last term represents the kinematical twist-three contribution from the bilocal operator. Its explicit expression is given later in Eq.~\eqref{eq:qg_transverse_ct}.

There is another possible twist-three correction originating from the slow \(x^+\)-dependence of  the target quark field.  This ``dynamical target'' term is distinct from the transverse-vector projection discussed above.  Expanding the background field as
\[
\bar\psi(x^+,x^-,x_\perp)
=
\bar\psi(y^+,x^-,x_\perp)
+
(x^+-y^+)\partial_+\bar\psi(y^+,x^-,x_\perp)
+\cdots,
\]
one finds that the term proportional to \(\partial_+\bar\psi\) cancels between the \(\bar\psi\)- and \(\psi\)-background amplitudes in both quark--gluon topologies.  Therefore,
\begin{align}
\left.|i\mathcal M_{qg}|^2\right|_{\rm dyn.\,target}
=0 .
\end{align}
The detailed derivation is given in Appendix~\ref{app:dyn_target_twobody}.

\subsection{ {Trilocal operators}  }

We now turn to the trilocal terms in
Eq.~\eqref{eq:cross_section_sector_decomposition}.  These terms are the parts of the cross section in which the hard scattering probes, in addition to the two background quark fields, one background field strength.  There are two sources of the trilocal terms.  In the quark--gluon channel the field strength arises from the first subleading term in one of the background-field building blocks.  In the quark--antiquark channel it arises from the interference of the two-quark background amplitude \(i\mathcal M_{q\bar q}^{(\psi\bar\psi)}\) with the single-\(F\) amplitude \(i\mathcal M_{q\bar q}^{(F)}\).

Let us first organize the quark--gluon channel using the building-block notation introduced in Sec.~\ref{sec:dijet-amplitudes}.  The leading amplitudes contain the products \(\mathcal K_0{\mathcal H}_0{\mathcal X}_0\) for topology \((a)\) and \(\mathcal K_0{\mathcal L}_0{\mathcal X}_0\) for topology \((b)\).  The trilocal amplitude is obtained by replacing one, and only one, of these leading blocks by its subleading single-\(F\) counterpart.  Equivalently, the field strength may come from the external quark line, the internal quark line, or the external gluon line.

We denote by \(i\mathcal M_{qg,X}^{(a,3)}\) the topology-\(a\) amplitude in which the block \(X\) supplies this field-strength insertion, with \(X={\mathcal X},\mathcal H,\mathcal K\).  The analogous notation for topology \((b)\) uses \(X={\mathcal X},\mathcal L,\mathcal K\).  The complete quark--gluon trilocal amplitude is then
\begin{align}
i\mathcal M_{qg}^{(3)}
=
\sum_{X={\mathcal X},\mathcal H,\mathcal K}
i\mathcal M_{qg,X}^{(a,3)}
+
\sum_{X={\mathcal X},{\mathcal L},\mathcal K}
i\mathcal M_{qg,X}^{(b,3)}
+
{\rm ch.c.},
\label{eq:Mqg_3body_sum}
\end{align}
where ``ch.c.'' denotes the charge-conjugate background-\(\psi\) branch, obtained with the rules of Appendix~\ref{app:charge_conjugation_rules}.  The individual terms in Eq.~\eqref{eq:Mqg_3body_sum} follow directly from the building-block rules in Appendix~\ref{app:qg_building_blocks}; no additional diagrammatic input is required.

Because \(i\mathcal M_{qg}^{(3)}\) already carries one field-strength insertion, it contributes at twist three through its interference with the bilocal quark--gluon amplitude,
\begin{align}
\left. d\sigma_{qg} \right|_{\rm triloc.}
\propto
2\,{\rm Re}\left[
\left(i\mathcal M_{qg}^{(3)}\right)
\left(i\mathcal M_{qg}^{(2)}\right)^\dagger
\right],
\label{eq:qg_3body_interference}
\end{align}
The quark--antiquark contribution is treated in the same operator basis below, but it enters through the mixed interference already indicated in Eq.~\eqref{eq:cross_section_sector_decomposition}, rather than through a separate square of \(i\mathcal M_{q\bar q}^{(F)}\).

The trilocal contribution is already of twist three, so no additional kinematic expansion of the hard part is required.  We now contract the color generator in the trilocal amplitude with the  generator in the lower-twist amplitude.  With
\({\rm tr}(t^a t^b)=\delta^{ab}/2\),
\begin{align}
\sum_a t^a t^a=C_F\,\mathbf 1,
\qquad
\sum_a t^a \bar F_{-i} t^a
=-\frac{1}{2N_c}\,\bar F_{-i}.
\label{eq:three_body_color_reduction}
\end{align}
In color space, the ordering of the generators fixes the overall color factor. The (Ft) ordering yields the usual factor \(C_F\), whereas the (tF) ordering, together with the corresponding (Fq)-ordered contribution, yields the subleading factor \(-1/(2N_c)\). After combining the charge-conjugate branch with the Hermitian-conjugate interference terms, the resulting trilocal structures can be classified by the coordinate assignment of the fermion and field-strength insertions and by their symmetry under interchange of the fermion endpoints.

For compactness, we introduce the three-slot operator
\begin{align}
\mathcal O_F^{\,i,[\pm]}[a,b,c]
\,\equiv\,&
\bar\Psi(a^-,a_\perp)
\bar F_{-i}(b^-,b_\perp)
\gamma^+\Psi(c^-,c_\perp)
\,\pm\,
\bar\Psi(c^-,c_\perp)
\bar F_{-i}(b^-,b_\perp)
\gamma^+\Psi(a^-,a_\perp).
\label{eq:generic_3body_operator}
\end{align}
The first and third slots specify the two fermion endpoints, while the second slot specifies the position of the field-strength insertion. In the quark--gluon channel, the assignments \([x,x',y]\) and \([x',x,y]\) distinguish the two possible orderings of the background-quark and field-strength insertions along the fermion line. The assignment \([x,y,x']\), on the other hand, arises from the quark--antiquark channel. The antisymmetric \([-]\) combination results from pairing the charge-conjugate branch with the Hermitian-conjugate interference, whereas the symmetric \([+]\) combination is associated with contributions containing the \(\sigma F\) structure in the amplitude.

\subsubsection{ {Longitudinally polarized photon}  }

For a longitudinally polarized virtual photon, the strict back-to-back limit gives a useful simplification of the trilocal interference.  The explicit \(\gamma^+\) trace shows that the channels \((b,3)\times(a,2)^\dagger\), \((a,3)\times(b,2)^\dagger\), and
\((b,3)\times(b,2)^\dagger\) do not contribute.  The only surviving trilocal coefficient is therefore the one from \((a,3)\times(a,2)^\dagger\), supplemented by the charge-conjugate branch and the Hermitian-conjugate interference.  After these pieces are combined into the reduced antisymmetric operator basis, the result is
\begin{align}
\left.
|i\mathcal M^L_{qg}|^2
\right|_{\rm triloc.}
&=
e^2 \int
dx^- d^2x_\perp\,
dy^- d^2y_\perp\,\int_{x^-}^{\infty}dx^{\prime -}
e^{i k_q^+ (x^- - y^-)
-i \Delta_{\perp}\cdot( x_\perp - y_\perp)}\na
&\times 2\,{\rm Re}\,\Bigg\{\,
\mathcal C_1^{L,i} 
 \,
\mathcal O_F^{\,i,[-]}[x,x',y]
+ \mathcal C_2^{L,i } 
 \,
\mathcal O_F^{\,i,[-]}[x',x,y]
\Bigg\}.
\label{eq:qg_l_cs}
\end{align}
The form above is just the simplified version of the previous channel-by-channel expression.  In particular, the charge-conjugate vector coefficient is equal to \(\mathcal C^{L}\) after the vector projection.  With \(\Delta_x^-\equiv x^{\prime -}-x^-\), the simplified  coefficient denotes 
\begin{align}
\mathcal C_1^{L,i }
&=
\frac{ iP^i\Delta_x^-\,\epsilon_f^2 z}{  (q^-)^2  {{\tilde P^4}} }
\left(
2zC_F+\frac{1}{N_c}
\right)
+\mathcal C_2^{L,i } ,
\nonumber\\
\mathcal C_2^{L,i }
&=
-\frac{2\, q^-  P^i  \,\epsilon_f^2 z^2\bar z}{ (q^-)^2   {{\tilde P^6}} }
\left(
2zC_F-\frac{1}{N_c}
\right).
\label{eq:qg_l_cs_coeffs}
\end{align}
 In addition, the evaluation of the quark--gluon trilocal contribution also produces a contact term.  The contact term is independent of \(x^{\prime -}\); it is therefore not a trilocal operator and is absorbed into the bilocal kinematical twist-3 contribution multiplying \(\mathcal O^+\). It gives
\begin{align}
\left.|i\mathcal M^L_{qg}|^2\right|_{\rm kin.3}^{(3\to2)}
&=
e^2 C_F \int
dx^- d^2x_\perp\,
dy^- d^2y_\perp\,
e^{i k_q^+ (x^- - y^-)
-i \Delta_{\perp}\cdot( x_\perp - y_\perp)}
\times 2\,{\rm Re}\,\Bigg\{\, \frac{4(\Delta \cdot P ) \,\epsilon_f^2 z^3\bar z}
{  q^-   {{\tilde P^6}}  } \, \mathcal O^{+} \Bigg\} .
\label{eq:qg_longitudinal_ct}
\end{align}

The quark--antiquark channel gives an additional genuine trilocal contribution to the longitudinal cross section.  In this channel the relevant color contraction differs from that in the quark--gluon channel: after the Fierz reduction, only the color-octet part contributes to the interference with the single-\(F\) amplitude, giving the color factor
\(-1/(2N_c)\).  The color-singlet part of the produced \(q\bar q\) pair drops out.  

After the longitudinal vector projection, the spin-insertion terms proportional to \(\mathbb S_1\) and \(\mathbb S_2\) do not contribute, since they contain \(\gamma^-\sigma^{-i}\) or \(\sigma^{-i}\gamma^-\) and are killed by \((\gamma^-)^2=0\).  The resulting longitudinal quark--antiquark trilocal contribution is
\begin{align}
\left.|i\mathcal M^L_{q\bar q}|^2\right|_{\rm triloc.}
&=
e^2 \int
dx^- d^2x_\perp\,
dy^- d^2y_\perp\,\int_{x^-}^{\infty}dx^{\prime -}
e^{i k_q^+ (x^- - y^-)
-i \Delta_{\perp}\cdot( x_\perp - y_\perp)} 2\,{\rm Re}\,\Bigg\{{ \!-} \frac{z\bar z \epsilon_f^2 }
{N_c\,k_q^+ {{\tilde P^4}}  (q^-)^2} P^i\,
\mathcal O_F^{\,i,[-]}[x,y,x'] \Bigg\}.
\label{eq:qqbar_l_cs}
\end{align} 

\subsubsection{ {Transversely polarized photon}  }

For a transversely polarized virtual photon, the trilocal interference terms are likewise projected onto the vector channel. The spin-insertion terms involving \(\sigma^{-i}\bar F_{-i}\) are absorbed into the hard coefficients and evaluated within the same \(\gamma^+\)-projected trace. Consequently, they do not require an independent tensor operator in the final operator basis.

After combining the charge-conjugate branch with the Hermitian-conjugate interference terms, the quark--gluon trilocal contribution takes the form
\begin{align}
\left.
|i\mathcal M^T_{qg}|^2
\right|_{\rm triloc.}
&=
e^2 \int
dx^- d^2x_\perp\,
dy^- d^2y_\perp\,\int_{x^-}^{\infty}dx^{\prime -}
e^{i k_q^+ (x^- - y^-)
-i \Delta_{\perp}\cdot( x_\perp - y_\perp)} \frac{1}{2(2q^-)^2}\na
&\times 2\,{\rm Re}\,\Bigg\{\,
\mathcal C_1^{T,i} 
 \,
\mathcal O_F^{\,i,[-]}[x,x',y]
+ \mathcal C_2^{T,i } 
 \,
\mathcal O_F^{\,i,[-]}[x',x,y] + \mathcal C_3^{T,i} 
 \,
\mathcal O_F^{\,i,[+]}[x,x',y]
+ \mathcal C_4^{T,i } 
 \,
\mathcal O_F^{\,i,[+]}[x',x,y]
\Bigg\}.
\label{eq:qg_T_3body}
\end{align}
The hard coefficients are decomposed according to
\begin{align}
\mathcal C_n^{T,i}(x,x')
=
P^i\left[
C_F\,  H_{n,F}^{T}(x,x')
-\frac{1}{2N_c} H_{n,A}^{T}(x,x')
\right]
+ E(x,x')\,P^i\, H_{n,E}^{T}(x,x') .
\label{eq:qg_T_3b_color_decomp}
\end{align}
Here,
\[
H_{n,F}^{T}=0 \quad \text{for } n=3,4,
\qquad
H_{n,E}^{T}=0 \quad \text{for } n=2,4.
\]
In particular, the symmetric ([+]) structures, which originate from the \(\sigma^{-i}\bar F_{-i}\) spin insertion, receive no contribution proportional to \(C_F\); their generator-ordering contribution is instead proportional to the subleading color factor \(-1/(2N_c)\). The corresponding hard functions are
\begin{align}
 H_{1,F}^{T}
&=
-\frac{4q^-z^2}{P_\perp^4\tilde P^6}
\Big[2P_\perp^6(z^2+\bar z^2)-P_\perp^4\tilde P^2z(z-\bar z)+3P_\perp^2\tilde P^4z(1+z)-\tilde P^6(3\bar z^2+2z)\Big]
+\frac{4iz\,\Delta_x^-}{P_\perp^2\tilde P^4\bar z}
\Big[P_\perp^4(z^2+\bar z^2)-\tilde P^4\bar z^2\Big],
\na[4pt]
  H_{1,A}^{T}
&=
-\frac{4q^-z}{P_\perp^4\tilde P^6}
\Big[2P_\perp^6(z^2+\bar z^2)-P_\perp^4\tilde P^2z^2-2P_\perp^2\tilde P^4z^2+2\tilde P^6(\bar z^2+z)\Big]
-\frac{4i\,\Delta_x^-}{P_\perp^2\tilde P^4\bar z}
\Big[P_\perp^4(z^2+\bar z^2)-\tilde P^4\bar z^2\Big],
\na[4pt]
 H_{1,E}^{T}
&=
\frac{2q^-z}{N_cP_\perp^4\tilde P^2}
\Big\{\Big[3P_\perp^2z^2-2\tilde P^2(\bar z^2+z)\Big]
+2N_cC_Fz\Big[3P_\perp^2z^2-\tilde P^2(3\bar z^2+2z)\Big]\Big\},
\na[6pt]
  H_{2,F}^{T}
&=
-\frac{4q^-z^2}{P_\perp^2\tilde P^6}
\Big[2P_\perp^4(z^2+\bar z^2)-P_\perp^2\tilde P^2z(z-\bar z)+2\tilde P^4z\Big],
\na[4pt]
 H_{2,A}^{T}
&=
-\frac{8q^-z}{\tilde P^6}
\Big[P_\perp^2(z^2+\bar z^2)-\tilde P^2z^2\Big]
+\frac{4iz\,\Delta_x^-}{P_\perp^2},
\na[6pt]
 H_{3,A}^{T}
&=
4q^-z\bar z^2\left[\frac{1}{P_\perp^4}-\frac{1}{\tilde P^4}\right], \qquad  H_{3,E}^{T}
=
4q^-z\bar z^2\,\frac{1 }{P_\perp^4}  ,
\na[4pt]
 H_{4,A}^{T}
&=
-4q^-z \left[\frac{(z^2+\bar z^2)}{\tilde P^4} -\frac{z^2}{\tilde P^2 P_\perp^2} \right].
\label{eq:qg_T_3b_coefficients}
\end{align}

After the \(\bar\psi\) and charge-conjugate \(\psi\) branches are combined, the two contact weights enter with opposite signs in the transverse \(\gamma^+\)-projected trace.  Thus
\begin{align}
  \left.|i\mathcal M^T_{qg}|^2\right|_{\rm kin.\,3}^{(3\to2)}&= e^2 C_F \int
dx^- d^2x_\perp\,
dy^- d^2y_\perp\,
e^{i k_q^+ (x^- - y^-)-i \Delta_{\perp}\cdot( x_\perp - y_\perp)}\na
&\times 2\,{\rm Re}\,\Bigg\{\,
\frac{z^2(\Delta\cdot P)}{{ (2q^-)}\tilde P^6 P_\perp^2}
\Big[ 2 P_\perp^4(z^2+\bar z^2) - P_\perp^2 \tilde{P}^2 \,z \,(z-\bar z) +2\, \tilde{P}^4 z \Big]\,
\mathcal O^{+}\Bigg\}.
\label{eq:qg_transverse_ct}
\end{align}

The quark--antiquark channel has the same color reduction as in the
longitudinal case.  The singlet term in the Fierz identity drops out against the traceless generator in the single-\(F\) amplitude, leaving the color factor \(-1/(2N_c)\).  For a transverse photon, the explicit \(\mathbb S_1\) and \(\mathbb S_2\) spin-insertion terms cancel in the complete trace.  The resulting contribution is
\begin{align}
\left.
|i\mathcal M^T_{q\bar q}|^2
\right|_{\rm triloc.}
&=
e^2 \int
dx^- d^2x_\perp\,
dy^- d^2y_\perp\,\int_{x^-}^{\infty}dx^{\prime -}
e^{i k_q^+ (x^- - y^-)
-i \Delta_{\perp}\cdot( x_\perp - y_\perp)}  2\,{\rm Re}\,\Bigg\{{ \!-}  \frac{1}{(2q^-)^2}
\frac{
 P_\perp^2(z-\bar z)
}{
N_c\,k_q^+ {{\tilde P^4}} 
}
P^i\,
\mathcal O_F^{\,i,[-]}[x,y,x'] \Bigg\}.
\label{eq:qqbar_T_cs}
\end{align}

\section{Conclusions}
\label{sec:conclusion}

In this work we computed the DIS dijet production cross section in the back-to-back limit, $\Delta_\perp \ll P_\perp$, at leading order in $\alpha_s$ in the presence of background quark fields.  We considered both longitudinal and transverse photon scattering on a hadronic target, and organized the cross section into hard coefficients multiplying quark TMD distributions. For the leading dynamical twist-two operator $\bar\psi\gamma^+\psi$, we included both the leading contribution and the subleading kinematic correction of order $O(\Delta_\perp/P_\perp)$. For the twist-three bilocal operator $\bar\psi\gamma^i\psi$ and the twist-three trilocal operator $\bar\psi\gamma^+F\psi$, we retain the leading dynamical twist contributions.

A central feature of our approach is that the background-field propagators are constructed from a transverse-gradient expansion.  This procedure keeps the full longitudinal phase $e^{i x P^+ z^-}$. The resulting expressions are valid at arbitrary Bjorken-$x$.  In the small-$x$ limit, our result reproduces the leading-twist quark operator $\bar\psi\gamma^+\psi$ of Ref.~\cite{Altinoluk:2023qfr}, obtained by computing the leading subeikonal correction to CGC.
In contrast to Ref.~\cite{Altinoluk:2023qfr}, our results extend to arbitrary Bjorken-$x$ and to the subleading twist-three contributions.  The latter constitute subsubeikonal corrections to CGC and are yet to be obtained from CGC.

Together with the contributions of gluon TMD distributions of Ref.~\cite{Mukherjee:2026cte}, the present results provide the complete leading-order in strong-coupling back-to-back dijet production cross section in DIS at arbitrary Bjorken-$x$ to twist-three accuracy. This theoretical framework constitute a necessary tools for phenomenological extractions of quark and gluon TMD distributions from future measurements at the EIC.


\acknowledgements

This work is supported by the U.S. Department of Energy, Office of Science, Office of Nuclear Physics through Contract Nos.~DE-SC0012704 and DE-SC0020081, the Saturated Glue (SURGE) Topical Collaboration in Nuclear Theory, and the United States-Israel Binational Science Foundation grant \#2022132. 

\newpage 

\appendix

\section{Back-to-back kinematics}
\label{app:btb_kinematics}

We follow the back-to-back kinematic conventions of
Ref.~\cite{Mukherjee:2026cte}, adapted here to the quark-background dijet
channels considered in this work.  The total transverse momentum imbalance and
the relative transverse momentum of the observed dijet are defined as
\begin{align}
\Delta_\perp
&\equiv
k_{1\perp}+k_{2\perp},
~~~~~~~~~~~~~~
P_\perp
\equiv
\bar z\,k_{1\perp}-z\,k_{2\perp},
\label{eq:btb_delta_P_def}
\end{align}
where
\begin{align}
z\equiv \frac{k_1^-}{q^-},
\qquad
\bar z\equiv \frac{k_2^-}{q^-}=1-z .
\label{eq:btb_z_def}
\end{align}
Equivalently,
\begin{align}
k_{1\perp}
&=
P_\perp+z\Delta_\perp,
~~~~~~~~~~~~~~
k_{2\perp}
=
-P_\perp+\bar z\Delta_\perp .
\label{eq:btb_inverse_transverse}
\end{align}
The back-to-back expansion is an expansion around
\(|\Delta_\perp|\ll |P_\perp|\).  The momentum transferred from the target
background to the hard subprocess is denoted throughout this work by
\begin{align}
k_q^\mu\equiv k_1^\mu+k_2^\mu-q^\mu .
\label{eq:kq_btb_def}
\end{align}
With \(q^\mu=(q^+,q^-,0_\perp)\) and \(q^+=-Q^2/(2q^-)\), the on-shell 
momenta are
\begin{align}
k_q^\mu
&=
\left(k_1^+ + k_2^+ - q^+,\,0,\,\Delta_\perp\right) 
 \approx
\left(
\frac{P_\perp^2+\epsilon_f^2}{2z\bar z\,q^-}
+\frac{\Delta_\perp^2}{2q^-},
\,0,\,
\Delta_\perp
\right),
\label{eq:kq_btb}
\\[4pt]
k_1^\mu
&=
\left(
\frac{(P_\perp+z\Delta_\perp)^2}{2zq^-},
\,zq^-,
\,P_\perp+z\Delta_\perp
\right)
 \approx
\left(
\frac{P_\perp^2}{2zq^-}
+\frac{\Delta \cdot P }{q^-}
+\frac{z\Delta_\perp^2}{2q^-},
\,zq^-,
\,P_\perp+z\Delta_\perp
\right),
\label{eq:k1_btb}
\\[4pt]
k_2^\mu
&=
\left(
\frac{(P_\perp-\bar z\Delta_\perp)^2}{2\bar z q^-},
\,\bar z q^-,
\,\bar z\Delta_\perp-P_\perp
\right)
 \approx
\left(
\frac{P_\perp^2}{2\bar z q^-}
-\frac{\Delta \cdot P }{q^-}
+\frac{\bar z\Delta_\perp^2}{2q^-},
\,\bar zq^-,
\,\bar z\Delta_\perp-P_\perp
\right).
\label{eq:k2_btb}
\end{align}
Here \(\epsilon_f^2\equiv z\bar z Q^2\).  The right hand sides of these equations represent the expansion in powers of \(\Delta_\perp/P_\perp\).

We reserve \(P^+\) for the plus momentum of the target hadron and \(P_\perp\)
for the relative transverse momentum defined in Eq.~\eqref{eq:btb_delta_P_def}.
Both are ordinary kinematic variables and should not be confused with the
covariant momentum operator \(\rP_\mu=p_\mu+A_\mu\). 

\section{Propagators}\label{app:propagators}
In this appendix, we derive the expressions for the quark and gluon propagators used in the main text. While the {quark} propagator was already computed up to twist 2 in Ref.~\cite{Mukherjee:2026cte}, it was computed under the assumption that the $p^-$ momentum in the propagator is positive. In this appendix, we generalize this result for any value of $p^-$ momenta up to twist 1. Going to twist 2 is a straightforward exercise following the procedure in 
Ref.~\cite{Mukherjee:2026cte}. Since both gluon and quark propagators contain parts that can be expressed in terms of the scalar propagator, let us first compute the general expression for the scalar propagator.
\subsection{Scalar propagator}

The scalar propagator can be simplified by using the expansion: 
\begin{align}
     (x| \frac{1}{\rP^2+ i \epsilon} |y)  &=
     \int \dhd p^- \int \dhd q^-  (x^+|p^-) \: (q^-|y^+)\:   (x^-,x_\perp| (p^-|  \frac{1}{{\rP}^2+ i \epsilon} |q^-)  |y^-,y_\perp) 
     \na
     &=  
     \int \frac{\dhd p^- }{2p^-} e^{-ip^-(x^+-y^+)}   (x^-,x_\perp| \frac{1}{{\rP^+} -  \frac{{\rP}_\perp^2}{2p^-} + i\,s_p\,  \epsilon }   |y^-,y_\perp)
     \na 
     & = 
      \int_{0}^\infty \frac{\dhd p^- }{2p^-} e^{-ip^-(x^+-y^+)}   (x^-,x_\perp| \frac{1}{{\rP^+} -  \frac{{\rP}_\perp^2}{2p^-} + i  \epsilon }   |y^-,y_\perp)
      \na& + 
      \int_{-\infty}^0 \frac{\dhd p^- }{2p^-} e^{-ip^-(x^+-y^+)}   (x^-,x_\perp| \frac{1}{{\rP^+} -  \frac{{\rP}_\perp^2}{2p^-} - i  \epsilon }   |y^-,y_\perp)\,.
\end{align}
We now perform a formal expansion in powers of
\(P_\perp^2/(2p^-)\).  For the positive \(p^-\) sector, this gives
\begin{align}
\label{eq:ScPr1}
    (x^-,x_\perp| \frac{1}{{\rP^+} -  \frac{{\rP}_\perp^2}{2p^-} + i  \epsilon }   |y^-,y_\perp) = 
    (x^-,x_\perp|  \frac{1}{{\rP^+}  + i  \epsilon }   |y^-,y_\perp)
    + (x^-,x_\perp|  \frac{1}{{\rP^+}  + i  \epsilon }   \frac{{\rP}_\perp^2}{2p^-} \frac{1}{{\rP^+}  + i  \epsilon }  |y^-,y_\perp) + \ldots 
\end{align}
Using the result reviewed in Appendix E.1 of Ref.~\cite{Mukherjee:2026cte},
one has
\begin{align}\label{eq:Pp_plus}
      (x^-|  \frac{1}{{\rP^+}  + i  \epsilon }   |y^-) = 
      - i \theta(x^--y^-)  [x^-, y^-]. 
\end{align}
The corresponding expression with negative $p^-$ can be obtained by Hermitian conjugation. Taking the Hermitian conjugate of Eq.~\eqref{eq:Pp_plus} gives
\begin{align}
\label{eq:Pp2Wl}
      (y^-|  \frac{1}{{\rP^+}  - i  \epsilon }   |x^-) = 
       i \theta(x^--y^-)  [y^-, x^-]. 
\end{align}
Combining these two equation together we get 
\begin{align}
\label{eq:Pp2Wlmp}
      (x^-|  \frac{1}{{\rP^+}  \pm i  \epsilon }   |y^-) = 
      \mp  i \theta(\pm(x^--y^-))  [x^-, y^-]. 
\end{align}
Returning back to Eq.~\eqref{eq:ScPr1}, 
we have 
\begin{align}
\label{eq:ScPr2}
    (x^-,x_\perp| \frac{1}{{\rP^+} -  \frac{{\rP}_\perp^2}{2p^-} + i  \epsilon }   |y^-,y_\perp) = 
    -i \theta(x^--y^-) (x_\perp|  [x^-,y^-] -i  \int_{y^-}^{x^-} d x'^- 
    [x^-,x'^-]  \frac{P_\perp^2(x'^-)}{2p^-} [x'^-,y^-]
    + \ldots |y_\perp).  
\end{align}
The structure of the higher-order terms is straightforward: each insertion of
\(P_\perp^2\) generates a transverse momentum kick. 
Since in the present analysis, we are interested only in the collinear
propagator up to the first subleading transverse-gradient correction, we
commute all transverse covariant derivatives through the Wilson lines to the left. This procedure was discussed in detail in Eqs.~(11)--(15) of
Ref.~\cite{Mukherjee:2026cte}. For the positive-\(p^-\) sector one obtains
\begin{align}
\label{eq:scalar_p_greater}
        (x^-,x_\perp| \frac{1}{{\rP^+} -  \frac{{\rP}_\perp^2}{2p^-} + i  \epsilon }   |y^-,y_\perp) \,&=\, \theta(x^- - y^-)\,(x_\perp| e^{ -i (x^--y^-) \frac{ {\rm P}_\perp^2 (x^-) }{2 p^-}}  \bigg[ -i \,\,[x^-, y^-]\,\na &+ \frac{1}{2p^-}  \int_{y^-}^{x^-} dx'^- (x'^- - y^-) \, 
    2 {\rm P}_k {(x^-)} \left[x^{-}, x'^{-} \right] F_{-k}(x'^-) [x'^-, y^-]\bigg]|y_\perp).
\end{align}
Here and in the following, \({\rm P}_k(x^-)\) denotes the transverse
covariant momentum after it was commuted to the left endpoint of the
Wilson line.  We  keep only the correction linear in \(F_{-k}\), which is
sufficient for twist-one accuracy in the transverse-gradient expansion.

We now derive the corresponding expression for the negative-\(p^-\) sector. Using Eq.~\eqref{eq:Pp2Wlmp}, the longitudinal propagation is reversed, and the support becomes \(y^->x^-\). At fixed negative \(p^-\), one finds
\begin{align}
\label{eq:scalar_p_smaller}
        (x^-,x_\perp| \frac{1}{{\rP^+} -  \frac{{\rP}_\perp^2}{2p^-} - i  \epsilon }   |y^-,y_\perp) \,&=\, - \theta(y^- - x^-)\,(x_\perp| e^{ -i (x^--y^-) \frac{ {\rm P}_\perp^2 (x^-) }{2 p^-}}  \bigg[ -i \,\,[x^-, y^-]\,\na &+ \frac{1}{2p^-}  \int_{y^-}^{x^-} dx'^- (x'^- - y^-) \, 
    2 {\rm P}_k (x^-) \left[x^{-}, x'^{-} \right] F_{-k}(x'^-) [x'^-,y^-]\bigg]|y_\perp)\,,
\end{align}
where the additional sign in the twist 1 piece comes from flipping the limits of the $x^-$ integral, which is naturally from $x^-$ to $y^-$ due to the theta function $\theta(y^--x^-)$. For an arbitrary sign of $p^-$, we can therefore write the general expression
\begin{align}
 (x| \frac{1}{{\rP^+} -  \frac{{\rP}_\perp^2}{2p^-} + i  \epsilon }   |y) \,&=\,\int_{-\infty}^{\infty}\,\frac{\dhd p^-}{2p^-}\,s_p\,\Theta_p(x^-,y^-)\,e^{-i p^-(x^+ - y^+)}\,(x_\perp| e^{ -i (x^--y^-) \frac{ {\rm P}_\perp^2 (x^-) }{2 p^-}}  \na &\times\bigg[ -i \,\,[x^-, y^-]\,+ \frac{1}{2p^-}  \int_{y^-}^{x^-} dx'^- (x'^- - y^-) \, 
    2 {\rm P}_k (x^-) \left[x^{-}, x'^{-} \right] F_{-k}(x'^-) [x'^-,y^-]\bigg]|y_\perp)\,,
\end{align}
where $s_p$ and $\Theta_p$ were defined in Eq.~\eqref{eq:sign_function}.

\subsection{Gluon propagator}
The gluon propagator in the background Feynman gauge is given by 
\begin{align}
&\mathcal G_{\mu \nu}(x,y)\:=\:(x|\frac{-i}{g \rP^2 + 2 i  F + i \epsilon}|y)_{\mu \nu} 
\:=\:(x| \frac{-i}{g \rP^2 + i \epsilon} - \frac{-i}{g \rP^2 + i \epsilon}\, 2 i F\,\frac{1}{g \rP^2 + i \epsilon}|y)_{\mu \nu}.
\end{align}
Using the identity
\begin{align}
g^{\mu \nu} g_{\nu \alpha}\:=\:\delta^{\mu}_{\alpha},
\end{align}
we obtain
\begin{align}
\label{eq:gluon_prop_before_expansion}
&G_{\mu \nu}(x,y)\:
\:=\:(x| \frac{-i\,g_{\mu \nu}}{{\rP}^2+ i \epsilon} - \frac{ -i } { \rP^2+ i \epsilon}\, 2i F_{\mu \nu} \frac{1} { \rP^2+ i \epsilon}|y).
\end{align}
We can now substitute the result of the previous subsection for the scalar propagator.
To twist--1 precision, the scalar propagator in the second term of Eq. \eqref{eq:gluon_prop_before_expansion}  can be simply replaced by the Wilson line, thus yielding:
\begin{align}
 G_{\mu \nu}(x,y)&= \!-i\int_{-\infty}^{\infty}\,\frac{\dhd p^-}{2p^-}\,s_p\,\Theta_p(x^-, y^-) { e^{-ip^-(x^+-y^+)}} (x_\perp| e^{ -i (x^--y^-) \frac{ {\rm P}_\perp^2 (x^-) }{2 { p^-}}}\bigg[ -i \,g_{\mu \nu} [x^-, y^-] + \frac{g_{\mu \nu}}{2p^-}  \int_{y^-}^{x^-} dx'^- (x'^- - y^-) \na & \times \, 
    2 {\rm P}_k (x^-) \left[x^{-}, x'^{-} \right] F_{-k}(x'^-)[x'^-,y^-]\: +\:\frac{i}{p^-} \int_{y^-}^{x^-}\,dx'^-\,[x^-,x'^-]F_{\mu\nu}(x'^-) [x'^-,y^-]\bigg]|y_\perp).    
\end{align}

\subsection{Quark propagator}
  In the background field, the quark propagator is given by
\begin{align}
\mathcal S_{\slashed{\rm {P}}}^{ij}(x,y)\,&=\,(x| \slashed{P}\frac{1}{\rP^2+ \frac{1}{2}\sigma F + i \epsilon} |y)^{ij}\na & =
     \int \dhd p^- \int \dhd q^-  (x^+|p^-) \: (q^-|y^+)\:   (x^-,x_\perp| (p^-| \slashed{P} \frac{1}{\rP^2+ \frac{1}{2}\sigma F + i \epsilon} |q^-)  |y^-,y_\perp)^{ij} \na &
     = \int_{-\infty}^{\infty} \dhd p^- \,e^{-i p^- (x^+ - y^+)} \,  (x^-,x_\perp| (\gamma^- {\rP^+} + \gamma^+ p^- - \gamma_\perp P_\perp) \frac{1}{2 p^- {\rP^+} - P_\perp^2+ \frac{1}{2}\sigma F + i \epsilon}  |y^-,y_\perp)^{ij}
\end{align}
As for the scalar propagator, we want to evaluate this expression starting with the integral over the expectation value of  ${\rP^+}$. This, however, will lead to a problem as the ratio falls as $1/{\rP^+}$ rendering the associated integral divergent.  To circumvent this problem, we isolate the ${\rP^+}$-independent part of the quark propagator~\footnote{In LCPT, the ${\rP^+}$-independent part corresponds to the instantaneous vertex.}  
\begin{align}
    &\left({\rP^+}\gamma^-+p^-\gamma^+-{\rm {P}_\perp}\gamma_\perp\right)
{1\over {\rm {P}^2}+{1\over 2}\sigma F+i\epsilon\,s_p} \na=&  \frac{1}{2 p^-}   \left({\rP^+}\gamma^-+p^-\gamma^+-{\rm {P}_\perp}\gamma_\perp\right)
{1\over {\rP^+}-{{\rm {P}_\perp^2}\over 2p^-} +{\sigma F\over 4p^-}+i\epsilon\,s_p} \notag \\ =&
\frac{\gamma^-}{2p^-}  + \frac{1}{2 p^-} \left[ \gamma^- \left( { {\rm {P}_\perp^2}\over 2p^-}
- {\sigma F\over 4p^-} \right) + p^-\gamma^+-{\rm {P}_\perp}\gamma_\perp  \right]{1\over {\rP^+}-{{\rm {P}_\perp^2}\over 2p^-} +{\sigma F\over 4p^-}+i\epsilon\,s_p}\,.
\end{align}
The first term is local both in \(x^-\) and \(x_\perp\)
The remaining term has a numerator independent of \(P^+\), thus we can apply the methods used for the scalar propagator to compute it.

Substituting this decomposition into the quark propagator, we obtain
\begin{align}
\mathcal S_{\slashed{\rm {P}}}^{ij}(x,y)  =\,& \int_{-\infty}^{\infty} \frac{\dhd p^-}{2p^-} \,e^{-i p^- (x^+ - y^+)} \bigg\{\gamma^-\,\delta^{ij}\,\delta^2(x_\perp - y_\perp)\,\delta(x^- - y^-) \,+ (x^-,x_\perp| \bigg[ \gamma^- \left( { {\rm {P}_\perp^2}\over 2p^-}
- {\sigma F\over 4p^-} \right)  + p^-\gamma^+-{\rm {P}_\perp}\gamma_\perp  \bigg] \:\na &
{1\over {\rP^+}-{{\rm {P}_\perp^2}\over 2p^-} +{\sigma F\over 4p^-}+i\epsilon\,s_p} |y^-,y_\perp)^{ij}\bigg\}
\end{align}
Now the last factor can be easily computed either directly or by using the analogy to the gluon propagator  
\begin{align}
(x^-,x_\perp|&
{1\over {\rP^+}-{{\rm {P}_\perp^2}\over 2p^-} +{\sigma F\over 4p^-}+i\epsilon\,s_p} |y^-,y_\perp)
= 
 \,s_p \,\Theta_p(x^-,y^-)\, e^{ -i (x^--y^-) \frac{ {\rm P}_\perp^2 (x^-) }{2 p^-}}  \bigg\{ -i \,\,[x^-, y^-]\,\na & + \frac{1}{2p^-}  \int_{y^-}^{x^-} dx'^-\,(x'^- - y^-)[x^-,x'^-]\left[2P_k(x'^-)F_{-k}(x'^-)+{\sigma F(x'^-)\over 2(x'^- - y^-)}\right][x'^-,y^-]\bigg\}\,.
\end{align}
Putting everything together, we obtain twist--1 expression for the quark propagator: 
\begin{align}
\mathcal S_{\slashed{\rm {P}}}^{ij}(x,y)
&=
\int_{-\infty}^{\infty}\frac{\dhd p^-}{2p^-}\,
e^{-i p^- (x^+-y^+)}
\Bigg\{
\gamma^-\,\delta^{ij}\delta^2(x_\perp-y_\perp)\delta(x^- - y^-)
\,+\,s_p\,\Theta_p(x^-,y^-)
\notag\\
&\quad\times
\left(x_\perp\left|
e^{-i(x^- - y^-)\frac{{\rm {P}}_\perp^2(x^-)}{2p^-}}
\mathcal N_q(x^-;p^-,{\rm {P}_i})\bigg\{
-i\,[x^-,y^-]
+
\frac{1}{2p^-}
\int_{y^-}^{x^-}dx'^-\,(x'^- - y^-)
\right.
\right.
\notag\\
&\qquad\qquad\left.
\left.
\times
[x^-,x'^-]
\left[
2P_k(x'^-)F_{-k}(x'^-)
+\frac{\sigma F(x'^-)}{2(x'^- - y^-)}
\right]
[x'^-,y^-]
\bigg\}
\right|y_\perp\right)^{ij}
\Bigg\}\,,
\end{align}
where
\begin{align}
\mathcal N_q(x^-;p^-, {\rm {P}_i})
\equiv\;&
\gamma^-
\left[
{{\rm {P}_\perp^2}(x^-)\over 2p^-}
-
{\sigma F(x^-)\over 4p^-}
\right]
+
p^-\gamma^+
-
\gamma_\perp {\rm {P}_\perp}(x^-).
\end{align}

\section{The amplitude of the quark--gluon channel}
\label{app:qg_building_blocks}

In this appendix we derive the background-field building blocks used in the quark--gluon channel.  We keep terms with at most one subleading insertion in the twist expansion, while the free transverse propagation is kept unexpanded.  The building blocks associated with the external lines are common to the two topologies, whereas the internal quark-line blocks are topology dependent.

\subsection{Useful commutator identities}

We first collect the commutator identities needed below. Throughout this appendix, all equalities are understood up to twist-two accuracy. We define the Wilson-line-dressed field strength by
\begin{align}
\bar F_{\alpha\beta}(\xi^-)
\equiv
[\infty,\xi^-]\,
F_{\alpha\beta}(\xi^-,x^+)\,
[\xi^-,\infty],
\end{align}
and the Wilson-line-dressed background antiquark field by
\begin{align}
\bar\Psi(x^-)
\equiv
\bar\psi(x)\,[x^-,\infty].
\end{align}
Commuting one transverse covariant momentum through the semi-infinite
Wilson line gives
\begin{align}
\label{eq:comm_Pi_W_app}
\big[{\rP}_i,[\infty,x^-]\big]
=
\int_{x^-}^{\infty} d\xi^-\,
\bar F_{-i}(\xi^-)\,
[\infty,x^-].
\end{align}
Therefore
\begin{align}
\label{eq:comm_Pperp_W_app}
\big[{\rP}_\perp^2,[\infty,x^-]\big]
&=
\left\{
{\rP}_i,
\big[{\rP}_i,[\infty,x^-]\big]
\right\}
=
2p_i
\int_{x^-}^{\infty} d\xi^-\,
\bar F_{-i}(\xi^-)\,
[\infty,x^-].
\end{align}
In last step \(P_i\) was replaced by
the ordinary transverse momentum \(p_i\).  If it is commuted further through
the field-strength insertion, one generates terms with two subleading
insertions, which are beyond the required accuracy.

Similarly,
\begin{align}
\label{eq:comm_Pmu_W_app}
\big[{\rP}_\mu,[\infty,x^-]\big]
&=
\frac{\delta_{\mu -}}{2p^-}
\big[{\rP}_\perp^2,[\infty,x^-]\big]
+
\delta_{\mu i}
\big[{\rP}_i,[\infty,x^-]\big]
\na
&=
\left(
\frac{p_i\delta_{\mu -}}{p^-}
+
\delta_{\mu i}
\right)
\int_{x^-}^{\infty} d\xi^-\,
\bar F_{-i}(\xi^-)\,
[\infty,x^-].
\end{align}
We also need the corresponding identities when the momentum is commuted
through the dressed background antiquark field.  To the same accuracy,
\begin{align}
\label{eq:comm_Pi_qbar_app}
\big[
p_i,
\bar\psi(x)\,[x^-,\infty]
\big]
&=
iD_i\bar\psi(x)\,[x^-,\infty]
+
\bar\Psi(x^-)
\int_{x^-}^{\infty} d\xi^-\,
\bar F_{-i}(\xi^-).
\end{align}
Consequently,
\begin{align}
\label{eq:comm_Pperp_qbar_app}
\big[
p_\perp^2,
\bar\psi(x)\,[x^-,\infty]
\big]
&=
\left\{
p_i,
\big[
p_i,
\bar\psi(x)\,[x^-,\infty]
\big]
\right\}
\na
&=
2p_i
\left[
iD_i\bar\psi(x)\,[x^-,\infty]
+
\bar\Psi(x^-)
\int_{x^-}^{\infty} d\xi^-\,
\bar F_{-i}(\xi^-)
\right],
\end{align}
and
\begin{align}
\label{eq:comm_Pmu_qbar_app}
\big[
p_\mu,
\bar\psi(x)\,[x^-,\infty]
\big]
&=
\frac{\delta_{\mu -}}{2p^-}
\big[
p_\perp^2,
\bar\psi(x)\,[x^-,\infty]
\big]
+
\delta_{\mu i}
\big[
p_i,
\bar\psi(x)\,[x^-,\infty]
\big]
\na
&=
\left(
\frac{p_i\delta_{\mu -}}{p^-}
+
\delta_{\mu i}
\right)
\left[
iD_i\bar\psi(x)\,[x^-,\infty]
+
\bar\Psi(x^-)
\int_{x^-}^{\infty} d\xi^-\,
\bar F_{-i}(\xi^-)
\right].
\end{align}

\subsection{Universal external-line building blocks}
\label{app:universal_external_blocks}

We first collect the external-line building blocks that are common to the two quark--gluon topologies considered in the main text.  These blocks are ``universal'' in the sense that their derivation does not depend on the topology-specific internal quark-line block. The latter will be denoted by \(
\mathcal H\) for topology (a) and by \(\mathcal L\) for topology (b).

Throughout this subsection we keep terms with at most one subleading insertion in the twist expansion, while the free transverse propagation is kept in its exponentiated form and is not expanded.  

The LSZ-reduced external gluon line, including the color generator attached to the quark line, is defined by
\begin{align}
\label{eq:J_gluon_def_app}
\mathcal J_{\lambda\nu}^{a}(k_2;x)_{ij}
\equiv
\lim_{k_2^2\to0}
k_2^2\,
\mathcal G_{\lambda\nu}^{ab}(k_2,x)\,
t^b_{ij}.
\end{align}
Using the adjoint-to-fundamental identity
\begin{align}
\label{eq:adjoint_fund_identity_app}
[\infty,x^-]_{A}^{bc}\,t^c
=
[x^-,\infty]\,
t^b\,
[\infty,x^-],
\end{align}
the LSZ-reduced gluon insertion can be written as
\begin{align}
\label{eq:J_gluon_K_app}
\mathcal J_{\lambda\nu}^{a}(k_2;x)_{ij}
=
e^{ik_2^+x^-+ik_2^-x^+}
\left(
k_{2\perp}
\left|
\mathcal K_{\lambda\nu}^{ac}(k_2;x^-)
\right|
b_\perp
\right)
\left(
[x^-,\infty]\,
t^c\,
[\infty,x^-]
\right)_{ij}.
\end{align}
Here the Wilson lines in the last factor are in the fundamental
representation, while the Wilson lines inside \(\mathcal K\) below are in the adjoint representation.

The external-gluon kernel is expanded as
\begin{align}
\label{eq:K_expansion_app}
\mathcal K_{\lambda\nu}^{ac}
=
\mathcal K_{0,\lambda\nu}^{ac}
+
\mathcal K_{1,\lambda\nu}^{ac}
+\cdots .
\end{align}
The leading-power contribution is
\begin{align}
\label{eq:K0_app}
\mathcal K_{0,\lambda\nu}^{ac}(k_2;x^-)
=
g_{\lambda\nu}\,\delta^{ac},
\end{align}
and the one-subleading-insertion contribution is
\begin{align}
\label{eq:K1_app}
\mathcal K_{1,\lambda\nu}^{ac}(k_2;x^-)
&=
\frac{i}{2k_2^-}
\int_{x^-}^{\infty} dz_1^-\,
(z_1^- - x^-)
\left[
[\infty,z_1^-]_{A}
\left(
2g_{\lambda\nu}\,k_{2r}F_{-r}(z_1)
+
\frac{2iF_{\lambda\nu}(z_1)}
{z_1^- - x^-}
\right)
[z_1^-,\infty]_{A}
\right]^{ac}.
\end{align}
In Eq.~\eqref{eq:K1_app}, the first term in the parentheses comes from the transverse-momentum insertion in the gluon propagation, while the second term comes from the local spin-dependent field-strength insertion.  Both terms contain one subleading background-field insertion and are therefore kept at the accuracy required for the twist-three amplitude.

The second universal external-line block is the amputated external fermion block.  It is obtained by combining the semi-infinite Wilson line ending at \(y^-\) with the LSZ-amputated external fermion propagator:
\begin{align}
\label{eq:A_block_def_app}
&[\infty,y^-]^{mk}
\lim_{k_1^2\to0}
k_1^2
\left(
y
\left|
\frac{i}
{{\rm P}^2+\frac12\sigma F+i\epsilon}
\right|
-k_1
\right)^{kl}
 =
e^{ik_1^+y^-+ik_1^-y^+-ik_{1\perp}\cdot y_\perp}
\left[
{\mathcal X}_0(k_1,y)+{\mathcal X}_1(k_1,y)
\right]^{ml}.
\end{align}
The leading-power block is simply
\begin{align}
\label{eq:A0_app}
{\mathcal X}_0(k_1,y)
=
1.
\end{align}
The contribution with one subleading insertion on the external fermion line is
\begin{align}
\label{eq:A1_app}
{\mathcal X}_1(k_1,y)
=
-\frac{i}{2k_1^-}
\int_{y^-}^{\infty} dz_1^-\,
(z_1^- - y^-)
\left[
2k_{1r}\,\bar F_{-r}(z_1^-)
-
\frac{\sigma\bar F(z_1^-)}
{2(z_1^- - y^-)}
\right],
\end{align}
where
\begin{align}
\sigma\bar F(z_1^-)
\equiv
\sigma^{\alpha\beta}\bar F_{\alpha\beta}(z_1^-),
\end{align}
and \(\bar F_{\alpha\beta}\) denotes the Wilson-line-dressed field strength in the fundamental representation.  Although we do not indicate explicitly, we imply that the background fields in Eq.~\eqref{eq:A1_app} are evaluated at  \(y_\perp\) and \(y^+\).

Equations~\eqref{eq:K_expansion_app}--\eqref{eq:A1_app} define the four external-line building blocks used in both topologies,
\begin{align}
\mathcal K_0,\qquad
\mathcal K_1,\qquad
{\mathcal X}_0,\qquad
{\mathcal X}_1 .
\end{align}
The topology dependence of the quark--gluon amplitude enters only through the internal quark-line block, which is \(\mathcal H\) for topology (a) and \(\mathcal L\) for topology (b).  To the accuracy considered here, products involving two subleading external-line insertions, such as \(\mathcal K_1 {\mathcal X}_1\), are dropped.

\subsection{Internal quark-line block for topology (a)}

We now apply these identities to the quark line. It is useful to introduce the notation for the operator in the numerator
\begin{align}
\label{eq:GammaP_def_app}
\Gamma_{\rP}(x^-,y^-)
\equiv
\left(
i\delta(x^- - y^-)
+
\frac{{\rP}_\perp^2}{2p^-}
\right)\gamma^-
+
p^-\gamma^+
+
{\rP}_\perp^r\gamma_r .
\end{align}
After the covariant momenta are commuted through the Wilson lines and the
background quark field, \(\Gamma_P\) reduces, at the order considered here,
to the ordinary momentum structure
\begin{align}
\label{eq:Gamma0_def_app}
\Gamma_0(x^-,y^-;p^-)
\equiv
\left(
i\delta(x^- - y^-)
+
\frac{p_\perp^2}{2p^-}
\right)\gamma^-
+
p^-\gamma^+
+
p_\perp^r\gamma_r .
\end{align}

We first combine quark propagator with the semi-infinite Wilson line attached to the LSZ-reduced external gluon.  This step dresses the internal quark line by the Wilson line \([\infty,x^-]\).
Since the transverse covariant momentum does not commute with the
semi-infinite Wilson line, this rearrangement generates additional
commutator terms.  Keeping terms up to twist-2 accuracy, these
non-commuting contributions can be absorbed into an effective numerator
factor \(\mathcal N(x^-,y^-)\), while the remaining Wilson-line propagation between \(y^-\) and \(x^-\), including one field-strength insertion inside the interval, is collected into \(\mathcal W(x^-,y^-)\).  In this way, the dressed quark-line factor takes the compact form
\begin{align}
\label{eq:dressed_prop_raw_app}
&\bar\psi^i(x)\,
\gamma^\nu
[x^-,\infty]^{im}t^c_{mn}[\infty,x^-]^{nj}
\mathcal S_{\slashed{\rm P}}^{jk}(x,y)\na
&=
\int_0^\infty
\frac{\dhd p^-}{2p^-}\,
e^{-ip^-(x^+-y^+)}
(
x_\perp
| \bar\psi(x)\gamma^\nu[x^-,\infty]t^c\,
e^{-i(x^- - y^-)\frac{p_\perp^2}{2p^-}}\,
\mathcal N(x^-,y^-)\,
\mathcal W(x^-,y^-)
 |
y_\perp
 )^{nk},
\end{align}
where
\begin{align}
\label{eq:N_def_app}
\mathcal N(x^-,y^-)= 
\Gamma_{0}(x^-,y^-)\,[\infty,x^-]
+
i\,\Gamma_{0}(x^-,y^-)
\frac{x^- - y^-}{2p^-}
\big[{\rP}_\perp^2,[\infty,x^-]\big]
-
\big[
\Gamma_{\rP}(x^-,y^-),[\infty,x^-]
\big]
-
[\infty,x^-]\,
\frac{\gamma^-\sigma F(x^-)}{4p^-},
\end{align}
and
\begin{align}
\label{eq:W_def_app}
\mathcal W(x^-,y^-)
=&\,
-i[x^-,y^-]
 +
\frac{1}{2p^-}
\int_{y^-}^{x^-} d\xi^-\,
(\xi^- - y^-)
[x^-,\xi^-]
\left[
2{\rP}_k F_{-k} (\xi^-)
+
\frac{\sigma F(\xi^-)}
{2(\xi^- - y^-)}
\right]
[\xi^-,y^-].
\end{align}

The product \(\mathcal N\mathcal W\) is expanded according to the number of subleading insertions.  We decompose
\begin{equation}
\mathcal N=\mathcal N_0+\mathcal N_1,\qquad
\mathcal W=\mathcal W_0+\mathcal W_1,
\end{equation}
where the subscript counts the number of transverse-derivative or
field-strength insertions beyond the eikonal Wilson-line propagation.  Up to twist-two accuracy, only terms with at most one such insertion are
kept.  Hence
\begin{equation}
\mathcal N\mathcal W
\simeq
\mathcal N_0\mathcal W_0
+
\mathcal N_1\mathcal W_0
+
\mathcal N_0\mathcal W_1,
\end{equation}
while \(\mathcal N_1\mathcal W_1\) is beyond the accuracy considered here. Accordingly, the leading block \({\mathcal H}_0\) is generated by
\(\mathcal N_0\mathcal W_0\), and the twist-two correction \({\mathcal H}_1\) by
\(\mathcal N_1\mathcal W_0+\mathcal N_0\mathcal W_1\).

Using Eqs.~\eqref{eq:comm_Pperp_W_app}--\eqref{eq:comm_Pmu_qbar_app}, and dropping higher-twist terms, the full expression up to twist-two becomes
\begin{align}
\label{eq:quark_line_full_tw3_app}
&\bar\psi^i(x)\,
\gamma^\nu
[x^-,\infty]^{im}t^c_{mn}[\infty,x^-]^{nj}
\mathcal S_{\slashed{\rm P}}^{jk}(x,y)\na [4pt]
&=\int_0^\infty
\frac{\dhd p^-}{2p^-}\,
e^{-ip^-(x^+-y^+)}
 (
x_\perp
 |
e^{-i(x^- - y^-)\frac{p_\perp^2}{2p^-}}
\bigg\{
-i\,
\bar\Psi(x^-)\,
\gamma^\nu
t^c
\Gamma_0  +
\bar\Psi(x^-)\,
\gamma^\nu
\Gamma_i
\bar F_{-i}^{(\infty)}(x^-)\,
t^c\na  [4pt]
&\quad+
\left(
iD_i\bar\psi(x)\,[x^-,\infty]
\right)
\gamma^\nu
\Gamma_i
t^c+
\bar\Psi(x^-)\,
\gamma^\nu
\Gamma_i
t^c\,
\bar F_{-i}^{(\infty)}(x^-) +
\frac{i}{4p^-}\,
\bar\Psi(x^-)\,
\gamma^\nu
t^c
\gamma^-\sigma\bar F(x^-)\na
&\quad +
\bar\Psi(x^-)\,
\gamma^\nu
t^c
\Gamma_0\,
\frac{1}{2p^-}
\int_{y^-}^{x^-} d\xi^-\,
(\xi^- - y^-)
\left[
2p_k\,\bar F_{-k}(\xi^-)
+
\frac{\sigma\bar F(\xi^-)}
{2(\xi^- - y^-)}
\right] \bigg\}
[\infty,y^-]
 |
y_\perp
 )^k,
\end{align}
where
\(\Gamma_0=\Gamma_0(x^-,y^-;p^-)\),
\(\Gamma_i=\Gamma_i(x^-,y^-;p^-)\), and we  introduced the shorthand notation
\begin{align}
\label{eq:F_infty_def_app}
\bar F_{-i}^{(\infty)}(x^-)
\equiv
\int_{x^-}^{\infty} d\xi^-\,
\bar F_{-i}(\xi^-),
\end{align}
and
\begin{align}
\label{eq:Gammai_def_app}
\Gamma_i(x^-,y^-;p^-)
\equiv
\frac{(x^- - y^-)\,p_i}{p^-}
\Gamma_0(x^-,y^-;p^-)
+
i\left(
\frac{p_i}{p^-}\gamma^-
+
\gamma_i
\right).
\end{align}
The two terms with \(\bar F_{-i}^{(\infty)}\) in Eq.~\eqref{eq:quark_line_full_tw3_app} have different color orderings.\\

We can now obtain the compact form used in the main text.  The
quark-line part is
\begin{align}
\label{eq:quark_line_H_app}
&\bar\psi^i(x)\,
\gamma^\nu
[x^-,\infty]^{im}t^c_{mn}
[\infty,x^-]^{nj}
\mathcal S_{\slashed{\rm P}}^{jk}(x,y)
\na[4pt]
&=
\int_0^\infty
\frac{\dhd p^-}{2p^-}\,
e^{-ip^-(x^+-y^+)}
 (
x_\perp
 |
e^{-i(x^- - y^-)\frac{p_\perp^2}{2p^-}}
\left[
  {\mathcal H}_0^{\nu c}(x^-,y^-; p)
+
  {\mathcal H}_1^{\nu c}(x^-,y^-; p)
\right]
[\infty,y^-]
 |
y_\perp
 )^k .
\end{align}
The leading block is
\begin{align}
\label{eq:H0_app}
  {\mathcal H}_0^{\nu c}(x^-,y^-; p)
=
-i\,
\bar\Psi(x^-)\,
\gamma^\nu
t^c
\Gamma_0(x^-,y^-;p^-).
\end{align}
The subleading block is
\begin{align}
\label{eq:H1_app}
  {\mathcal H}_1^{\nu c}(x^-,y^-; p)
=&\,
\left[
iD_i\bar\psi(x)\,[x^-,\infty]
\right]
\gamma^\nu
\Gamma_i
t^c
+
\bar\Psi(x^-)\,
\gamma^\nu
\Gamma_i
\bar F_{-i}^{(\infty)}(x^-)
t^c
+
\bar\Psi(x^-)\,
\gamma^\nu
\Gamma_i
t^c
\bar F_{-i}^{(\infty)}(x^-)+
\frac{i}{4p^-}\,
\bar\Psi(x^-)\,
\gamma^\nu
t^c
\gamma^-\sigma\bar F(x^-)
\na
&+
\bar\Psi(x^-)\,
\gamma^\nu
t^c
\Gamma_0\,
\frac{1}{2p^-}
\int_{y^-}^{x^-} d\xi^-\,
(\xi^- - y^-)
\left[
2p_k\,\bar F_{-k}(\xi^-)
+
\frac{\sigma\bar F(\xi^-)}
{2(\xi^- - y^-)}
\right],
\end{align}
where each term in \(  {\mathcal H}_1^{\nu c}\) contains exactly one subleading insertion: either \(D_i\bar\psi\), an integrated \(\bar F_{-i}\) from a semi-infinite Wilson line, an internal field-strength insertion from the propagator, or the local spin insertion \(\sigma\bar F\).  Terms with two or more such insertions are
dropped.

\subsection{Internal quark-line block for topology (b)}
\label{app:internal_quark_line_topology_b}

We now derive the internal quark-line block \(L\) used in
topology~(b).  The external-line building blocks \(\mathcal K\) and \({\mathcal X}\)
are the same as those derived above.  Thus the only topology-dependent
object is the quark-line factor
\begin{align}
\label{eq:L_quark_line_factor_app}
\mathcal S_{\slashed{\rm P}}^{ij}(y,x)\,
[x^-,\infty]^{jm}_{x_\perp}.
\end{align}
We focus on the ordering \(x^->y^-\), which is the ordering entering
Eq.~\eqref{eq:qg_b_barpsi}.  To the accuracy needed for the twist-three
amplitude, this factor can be written as
\begin{align}
\label{eq:L_block_def_app}
&
\mathcal S_{\slashed{\rm P}}^{ij}(y,x)\,
[x^-,\infty]^{jm}_{x_\perp}
\nonumber\\
&=
\int_{-\infty}^{0}
\frac{\dhd p^-}{2p^-}\,
e^{-ip^-(y^+-x^+)}
\left(
y_\perp
\left|
i\,[y^-,\infty]\,
\left[
{\mathcal L}_0(x^-,y^-;p)
+
{\mathcal L}_1(x^-,y^-;p)
\right]\,
e^{-i(y^- - x^-)\frac{p_\perp^2}{2p^-}}
\right|
x_\perp
\right)^{im}.
\end{align}
The integration range \(p^-<0\) is due to the reversed propagation of the
internal quark line in topology~(b).   The block \({\mathcal L}_0\) contains no subleading
background-field insertion, while \({\mathcal L}_1\) contains exactly one such insertion.

The leading block is
\begin{align}
\label{eq:L0_app}
  {\mathcal L}_0(x^-,y^-;p)
=
T_0^*(x^-,y^-;p),
\end{align}
where
\begin{align}
\label{eq:T0_star_app}
T_0^*(x^-,y^-;p)
\equiv
\left(
-i\delta(x^- - y^-)
+
\frac{p_\perp^2}{2p^-}
\right)\gamma^-
+
p^-\gamma^+
+
p_\perp^r\gamma_r .
\end{align}
The sign of the contact term is opposite to the corresponding topology-(a)
block because the propagator is written with the reversed ordering
\((y,x)\).

The subleading block receives three contributions.  The first comes from
commuting the transverse momentum through the semi-infinite Wilson line
\([x^-,\infty]\), the second is the local spin insertion at \(x^-\), and the
third comes from a single field-strength insertion inside the finite
propagation interval \([y^-,x^-]\).  Using the commutator identities collected
above, one obtains
\begin{align}
\label{eq:L1_app}
  {\mathcal L}_1(x^-,y^-;p)
&=
-i
\int_{x^-}^{\infty} d\xi^-\,
\bar F_{-i}(\xi^-)
\left[
\frac{(x^- - y^-)\,p_i}{p^-}\,
\slashed p
-
i\left(
\frac{p_i}{p^-}\gamma^-
+
\gamma_i
\right)
\right] 
-
\frac{\sigma\bar F(x^-)\gamma^-}{4p^-}
\nonumber\\
&\quad
-
i\,
\frac{1}{2p^-}
\int_{y^-}^{x^-} d\xi^-\,
(\xi^- - y^-)
\left[
2p_k\,\bar F_{-k}(\xi^-)
+
\frac{\sigma\bar F(\xi^-)}
{2(\xi^- - y^-)}
\right]\,
T_0^*(x^-,y^-;p).
\end{align}
Here
\begin{align}
\label{eq:slash_p_app}
\slashed p
\equiv
\frac{p_\perp^2}{2p^-}\gamma^-
+
p^-\gamma^+
+
p_\perp^r\gamma_r ,
\end{align}
namely the regular, non-contact part of \(T_0^*\).  In the first line of
Eq.~\eqref{eq:L1_app}, the contact term does not contribute because it is
multiplied by \((x^- - y^-)\delta(x^- - y^-)\).

Equations~\eqref{eq:L_block_def_app}--\eqref{eq:L1_app} define the
topology-(b) internal quark-line block used in Eq.~\eqref{eq:qg_b_barpsi}.
In the final compact amplitude the momentum flowing through this block is
fixed by the \(x^+\) and \(p^-\) integrations to
\begin{align}
p=-(k_1+k_2).
\end{align}
Terms containing two or more subleading insertions, such as a product of the
semi-infinite Wilson-line insertion with the finite-interval field-strength
insertion, are beyond the twist-three accuracy and were dropped.

\section{Charge-conjugation rules for the background-quark amplitudes}
\label{app:charge_conjugation_rules}

Here we list the rules to perform the charge conjugation.  For a fermion-line string \(\Gamma\), with transposition acting on both Dirac and fundamental color indices,
\begin{align}
\bar\psi\,\Gamma\,v(k)
\overset{C}{\longrightarrow}
-\,\bar u(k)\,
\left(C\Gamma^T C^{-1}\right)
\psi,
\end{align}
where \(C=i\gamma^2\gamma^0\).  Therefore the order of all factors on the fermion line is reversed.  In the notation of the main text this gives
\begin{align}
\bar\Psi(x^-)=\bar\psi(x)[x^-,\infty]
\quad\longrightarrow\quad
\Psi(x^-)=[\infty,x^-]\psi(x),
\end{align}
and the blocks \({\mathcal X}_i,{\mathcal H}_i,{\mathcal L}_i,\mathcal K\) are replaced by the corresponding tilded blocks with reversed fermion-line ordering.

For the background insertions used in this work, the required color-space rules are
\begin{align}
\big([\infty,x^-]F_{-i}(x^-)[x^-,\infty]\big)^{kl}
&\overset{C}{\longrightarrow}
-\big([\infty,x^-]F_{-i}(x^-)[x^-,\infty]\big)^{lk},
\na
\big([\infty,x^-]\sigma F(x^-)[x^-,\infty]\big)^{kl}
&\overset{C}{\longrightarrow}
\big([\infty,x^-]\sigma F(x^-)[x^-,\infty]\big)^{lk}.
\end{align}
The first minus sign follows from \(F_{\mu\nu}\to -F_{\mu\nu}^T\).  In the second line this minus sign is cancelled by the Dirac transformation of \(\sigma^{\mu\nu}\).

\section{Dynamical target bilocal term}
\label{app:dyn_target_twobody}

In this appendix we derive the bilocal contribution generated by the
\(x^+\)-dependence of the target quark field.  This term is separate from the dynamical twist-3 projection of the bilocal operator used in the main text. For the \(\bar\psi\)-background amplitude in the first quark--gluon topology, we expand
\begin{align}
\bar\psi(x^+,x^-,x_\perp)
&=
\bar\psi(y^+,x^-,x_\perp)
+
(x^+-y^+)\,
\partial_+\bar\psi(y^+,x^-,x_\perp)
+\cdots .
\label{eq:dyn_target_expansion}
\end{align}
Keeping the second term gives
\begin{align}
i\mathcal M_{qg,\bar\psi}^{(a,{\rm dyn.\,tar.})}
&=
-e\lim_{k^2,k_1^2\to0}
\int d^{\,4}x\,d^4y\,
e^{-iq\cdot y}\,
\epsilon_i^\lambda(k)\,
\epsilon_\mu(q)\,
k^2
\left(
k\left|
\frac{1}{P^2g_{\lambda\nu}+2iF_{\lambda\nu}+i\epsilon}
\right|z
\right)^{ij}
\na
&\quad\times
(x^+-y^+)
\partial_+\bar\psi^c(x^-,y^+,x_\perp)\,
t^j_{cb}\,
\gamma^\nu
\left(
z\left|
\frac{i}{P^2+\frac12\sigma F+i\epsilon}\slashed{P}
\right|y
\right)^{ba}
\gamma^\mu
\na
&\quad\times
\left(
y\left|
\frac{i}{P^2+\frac12\sigma F+i\epsilon}
\right|k_1
\right)^{ad}
k_1^2
v^d(k_1).
\label{eq:dyn_target_start}
\end{align}
The leading external gluon and quark LSZ reductions are the same as those used in Sec.~\ref{quark_bg:amp}.  The only new ingredient is the factor \((x^+-y^+)\) in the internal quark line.  In the instantaneous part of the quark propagator it produces the principal-value integral
\begin{align}
\mathcal I(k^-)
&\equiv
\int dx^+\,
e^{ik^-x^+}
\int_{-\infty}^{\infty}
\frac{\dhd p^-}{2p^-}\,
(x^+-y^+)\,
e^{-ip^-(x^+-y^+)}
\na
&=
-\frac{i\pi}{2}
\int dx^+\,
e^{ik^-x^+}
|x^+-y^+|
\na
&=
\frac{i\pi}{(k^-)^2}\,
e^{ik^-y^+},
\qquad k^->0,
\label{eq:dyn_target_pv_integral}
\end{align}
 The remaining non-instantaneous part can be treated by expressing 
\((x^+-y^+)e^{-ip^-(x^+-y^+)}=i\,\partial_{p^-}
e^{-ip^-(x^+-y^+)}\) and integrating by parts in \(p^-\).  After the
\(x^+\) integration, the result is 
\begin{align}
\frac{
i\mathcal M_{qg,\bar\psi}^{(a,{\rm dyn.\,tar.})}
}{
2\pi\delta(k^-+k_1^- -q^-)
}
&=
-ie
\int dx^-\,d^2x_\perp\,dy^-\,
e^{ik_q^+y^-}
e^{-ik_{q\perp}\cdot x_\perp}
\epsilon_i^\lambda(k)
\epsilon_\mu(q)
g_{\nu\lambda}
\na
&\quad\times
\partial_+\bar\psi^c(x^-,y^+,x_\perp)
[x^-,\infty]^{ce}
t^i_{eb}
\gamma^\nu
\mathcal R(k;x^-,y^-)
\gamma^\mu
v^b(k_1),
\label{eq:dyn_target_q_result}
\end{align}
with
\begin{align}
\mathcal R(k;x^-,y^-)
&=
\left[
-\frac{2\pi\,\delta(x^- - y^-)}{(k^-)^2}
+
\frac{k_\perp^2}{2k_q^+(k^-)^3}
-
\frac{(k_\perp^2)^2}{4(k_q^+)^2(k^-)^3}
\right]\gamma^-
\na
&\quad
-
\frac{k_\perp^2}{4(k_q^+)^2(k^-)^2}\,
\gamma^+
-
\left[
\frac{1}{2k_q^+(k^-)^2}
-
\frac{k_\perp^2}{4(k_q^+)^2(k^-)^3}
\right]
k_i\gamma^i .
\label{eq:dyn_target_R}
\end{align}
The detailed form of \(\mathcal R\) is irrelevant for the cancellation; the important point is that the overall factor \(-i\) in Eq.~\eqref{eq:dyn_target_q_result}. It is generated by the light-front time integral \eqref{eq:dyn_target_pv_integral}. The derivation above was written for the first quark--gluon topology.  For the second topology, the same \(x^+\)-expansion of the background quark field and the same light-front time integral appear in the corresponding internal line.   The order of the Dirac and color indices differ, but the relative phase between the \(\bar\psi\)- and \(\psi\)-background amplitudes is unchanged.

It is useful to isolate the phase structure by writing
\begin{align}
i\mathcal M_{qg,\bar\psi}^{({\rm dyn.\,tar.})}
&=
i A_{\bar\psi},\qquad
i\mathcal M_{qg,\psi}^{({\rm dyn.\,tar.})}
=
-i A_{\psi}, \qquad
i\mathcal M_{qg,\bar\psi}^{(2)}
=
B_{\bar\psi},\qquad
i\mathcal M_{qg,\psi}^{(2)}
=
-B_{\psi}.
\label{eq:dyn_target_phase_structure}
\end{align}
Here \(A_{\psi}\) and \(B_{\psi}\) are obtained from \(A_{\bar\psi}\) and \(B_{\bar\psi}\) by charge conjugation, with the same fermion-line ordering convention as in the bilocal amplitudes in
Sec.~\ref{sec:cross_section_operator_structure}. The contribution of the dynamical target term to the bilocal cross section is the interference with the lower-twist bilocal amplitude:
\begin{align}
\left.
|i\mathcal M_{qg}|^2
\right|_{\rm dyn.\,tar.}
&=
iA_{\bar\psi}B_{\bar\psi}^{\dagger}
+
(-iA_{\psi})(-B_{\psi})^{\dagger}
+
B_{\bar\psi}(iA_{\bar\psi})^{\dagger}
+
(-B_{\psi})(-iA_{\psi})^{\dagger}
\na
&=0 .
\label{eq:dyn_target_cancellation}
\end{align}
Thus the dynamical target bilocal correction does not contribute to the cross section at the accuracy considered in this paper.

\section{Trilocal hard coefficients}
\label{app:three_body_coefficients}

This appendix collects the explicit coefficient functions entering the compact trilocal amplitudes used in Secs.~\ref{sec:dijet-amplitudes} and~\ref{sec:cross_section_operator_structure}.  The formulas are organized by topology and by the origin of the field-strength insertion. To match the bilocal notation in the main text, coefficients in the first topology are denoted by \(\mathcal B\), while those in the second topology are denoted by \(\mathcal D\).  The superscript labels the source of the field-strength insertion; an additional superscript \(E\) identifies the coefficient of the term
containing the extra exponential factor. The first subscript, \(qF\) or \(Fq\), specifies the ordering of the background quark field \(\psi\) and the field-strength insertion \(F\) along the fermion propagator at the amplitude level.  This distinction
is not retained in the organization of the cross section. The second label indicates the color ordering: \(tF\) for \(\bar\psi t^a\bar F\), \(Ft\) for \(\bar\psi\bar F t^a\), \({\rm loc}\) for local field-strength terms, and \({\rm ct}\) for contact terms with bilocal operators. When used in the cross-section calculation, each amplitude listed below should be divided by the common normalization factor
\( 2\pi (2q^-) \delta \left(k_q^-\right) \). Throughout this appendix, the approximate coefficients are evaluated at leading order in the back-to-back expansion, so terms suppressed by powers of \(\Delta_\perp/P_\perp\) are dropped.

\subsection{First quark--gluon topology}

\begin{figure*}[t]
\centering
\begin{tabular}{@{}c@{\hspace{0.08\textwidth}}c@{\hspace{0.08\textwidth}}c@{}}
\begin{overpic}[
    width=0.25\textwidth
]{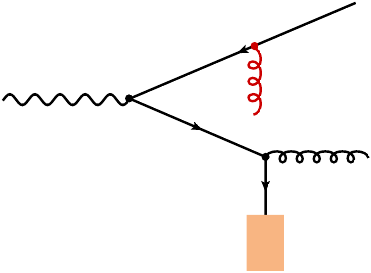}
    \put(3,54){\scriptsize $ {\gamma^*\,(q)}$}
    \put(98,71){\scriptsize $ {k_1}$}
    \put(101,31){\scriptsize $ {k_2}$}
    \put(70,40){\scriptsize $ \color{myred}{F}$}
    \put(69,6){\scriptsize $\boldsymbol{\bar\psi}$}
\end{overpic}
&
\begin{overpic}[
    width=0.25\textwidth
]{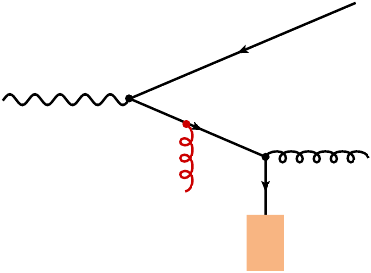}
    \put(3,54){\scriptsize $ {\gamma^*\,(q)}$}
    \put(98,71){\scriptsize $ {k_1}$}
    \put(101,31){\scriptsize $ {k_2}$}
    \put(48,16){\scriptsize $ \color{myred}{F}$}
    \put(69,6){\scriptsize $\boldsymbol{\bar\psi}$}
\end{overpic}
&
\begin{overpic}[
    width=0.25\textwidth
]{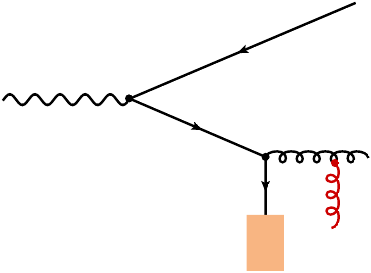}
    \put(3,54){\scriptsize $ {\gamma^*\,(q)}$}
    \put(98,71){\scriptsize $ {k_1}$}
    \put(101,31){\scriptsize $ {k_2}$}
    \put(87,7){\scriptsize $ \color{myred}{F}$}
    \put(69,6){\scriptsize $\boldsymbol{\bar\psi}$}
\end{overpic}
\\[4pt]
{\small (a1) $({\mathcal X}_1,\,{\mathcal H}_0,\,\mathcal K_0)$}
&
{\small (a2) $({\mathcal X}_0,\,{\mathcal H}_1,\,\mathcal K_0)$}
&
{\small (a3) $({\mathcal X}_0,\,{\mathcal H}_0,\mathcal K_1)$}
\end{tabular}
\caption{
Field-strength insertion points in the trilocal quark--gluon amplitude for
topology \((a)\).
}
\label{fig:qg_3body_topology_a}
\end{figure*}
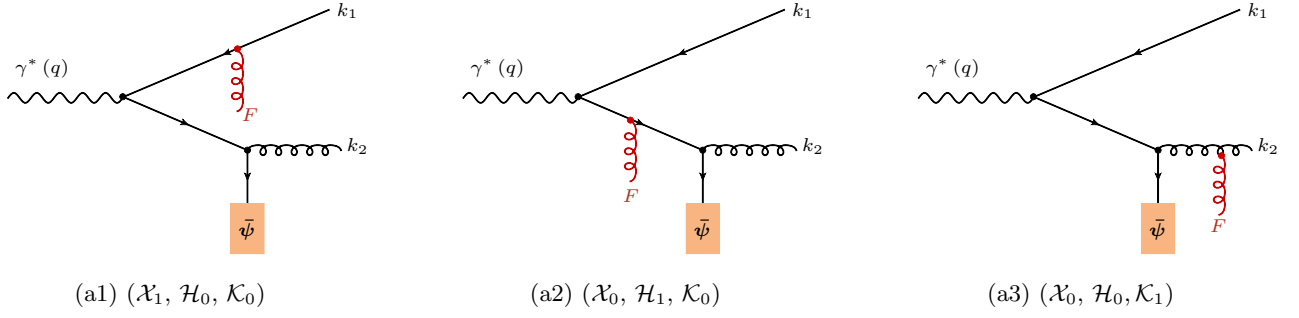

The three possible field-strength insertions in topology \((a)\) are shown in Fig.~\ref{fig:qg_3body_topology_a}.  They correspond to the subleading external quark block, internal quark-line block, and external gluon block, respectively.

The contribution from the LSZ-reduced external quark line is
\begin{align}
i\mathcal M_{qg,{\mathcal X}}^{(a,3)}
&=
-e
\int d^{\,4}x
\int_{x^-}^{\infty} dx^{\prime -}\,
 {e^{i k_q \cdot x } }
\epsilon_a^\lambda(k_2)\,
\epsilon_\mu(q)
\Big[
\big( \bar\Psi(x)\,t^a\,\bar F_{-r}(x')\big)^l\,
\mathcal B^{({\mathcal X})}_{qF,tF}(x,x')
+
\big(\bar\Psi(x')\,t^a\,\bar F_{-r}(x)\big)^l\,
\mathcal B^{({\mathcal X})}_{Fq,tF}(x,x')
\Big]v^l(k_1).
\label{eq:Mqg_a_A}
\end{align}
where
\begin{align}
\mathcal B^{({\mathcal X})}_{qF,tF}(x,x')
&=
\frac{\gamma_\lambda
\gamma^\nu
\gamma^\mu}{4k_1^- k_2^-}
\Bigg[
2 i k_{1r}\,\delta_\nu^-
(x^{\prime -}-x^-)
-i\delta_\nu^- \sigma^{-r}
+2 k_{1r} k_{2\nu}
\left(
\frac{i(x^{ -}-x'^-)}{k_q^+}
-\frac{1}{(k_q^+)^2}
\right)
+\frac{i k_{2\nu}}{k_q^+}\sigma^{-r}
\Bigg]\na
& \approx
\frac{\gamma_\lambda\gamma^\nu\gamma^\mu}
{4z\bar z\,(q^-)^2}
\Bigg[
2i\,P_r\delta_\nu^-(x^{\prime -}-x^-)
-i\delta_\nu^-\sigma^{-r}
+2P_r
\left(
\bar z q^-\delta_\nu^+
+ \frac{P_\perp^2}{2\bar z q^-}\delta_\nu^-
+ P_i\delta_\nu^i
\right)
\na
&\qquad\quad\times
\left(
\frac{2iz\bar z\,q^-(x^{ -}-x'^-)}
{ {{\tilde P^2}} }
-
\frac{4z^2\bar z^2(q^-)^2}
{ {{\tilde P^4}} }
\right)
+
\frac{2iz\bar z\,q^-}{ {{\tilde P^2}} }
\left(
\bar z q^-\delta_\nu^+
+ \frac{P_\perp^2}{2\bar z q^-}\delta_\nu^-
+ P_i\delta_\nu^i
\right)\sigma^{-r}
\Bigg]
\\[6pt]
\mathcal B^{({\mathcal X})}_{Fq,tF}(x,x')
&=
\frac{\gamma_\lambda
\gamma^\nu
\gamma^\mu}{4k_1^- k_2^-}
\Bigg[
2 i k_{1r}\,\delta_\nu^-
(x^- - x^{\prime -})
-i\delta_\nu^- \sigma^{-r} 
-2\frac{k_{1r} k_{2\nu}}{(k_q^+)^2}
+\frac{i k_{2\nu}}{k_q^+}\sigma^{-r}
\Bigg]\na [4pt]
& \approx 
\frac{\gamma_\lambda\gamma^\nu\gamma^\mu}
{4z\bar z\,(q^-)^2}
\Bigg[
2i\,P_r\delta_\nu^-(x^- - x^{\prime -})
-i\delta_\nu^-\sigma^{-r}
-
\frac{8z^2\bar z^2(q^-)^2}{ {{\tilde P^4}} }
P_r
\left(
\bar z q^-\delta_\nu^+
+ \frac{P_\perp^2}{2\bar z q^-}\delta_\nu^-
+ P_i\delta_\nu^i
\right)
\na
&\qquad
+
\frac{2iz\bar z\,q^-}{ {{\tilde P^2}} }
\left(
\bar z q^-\delta_\nu^+
+ \frac{P_\perp^2}{2\bar z q^-}\delta_\nu^-
+ P_i\delta_\nu^i
\right)\sigma^{-r}
\Bigg]
\label{eq:CA_coefficients}
\end{align}

The contribution from the subleading term in the internal quark propagator is
\begin{align}
i\mathcal M_{qg,\mathcal  H}^{(a,3)}
&=
-e 
\int d^{\,4}x\,
e^{i k_q\cdot x}\,
\epsilon_a^\lambda(k_2)\,
\epsilon_\mu(q)
\Bigg\{
\int_{x^-}^{\infty} dx^{\prime -}
\Big[
\big(\bar\Psi(x)\,t^a\,\bar F_{-r}(x')\big)^l
\,\mathcal B^{(\mathcal H)}_{qF,tF}(x,x')
+
\big(\bar\Psi(x)\,\bar F_{-r}(x')\,t^a\big)^l
\,\mathcal B^{(\mathcal H)}_{qF,Ft}(x,x')
\na
& \quad +
\big(\bar\Psi(x')\,t^a\,\bar F_{-r}(x)\big)^l
\,\mathcal B^{(\mathcal H)}_{Fq,tF}(x,x')
+
\big(\bar\Psi(x')\,\bar F_{-r}(x)\,t^a\big)^l
\,\mathcal B^{(\mathcal H)}_{Fq,Ft}(x,x')
\Big] 
+
\Big[
\big(\bar\Psi(x)\,t^a\,\bar F_{-r}(x)\big)^l
\,\mathcal B^{(\mathcal H)}_{\rm loc}
+
\big(\bar\Psi(x)\,t^a\big)^l
\,\mathcal B^{(\mathcal H)}_{\rm ct}
\Big]
\Bigg\}
v^l(k_1).
\label{eq:Mqg_a_H}
\end{align}
where  
\begin{align}
\mathcal B^{(\mathcal H)}_{qF,tF}(x,x')
&=
\frac{1}{2k_2^-k_q^+}\,
\gamma_\lambda\gamma^\eta
\left[
\frac{2k_2^r k_{2\eta}}{2k_2^-k_q^+}
-
\frac{2k_2^r g_{\eta}^{\ -}}{2k_2^-}
-
g_{\eta}^{\ r}
\right]\gamma^\mu
\na[4pt]
&\approx
\frac{z}{ {{\tilde P^2}} }\,
\gamma_\lambda\gamma^\eta
\left[
-\frac{2zP_r}{ {{\tilde P^2}} }
\left(
\bar z q^-\delta_\eta^+
+\frac{P_\perp^2}{2\bar z q^-}\delta_\eta^-
+P_i\delta_\eta^i
\right)
+
\frac{P_r}{\bar z q^-}g_\eta^{\ -}
-
g_\eta^{\ r}
\right]\gamma^\mu,
\na
\mathcal B^{(\mathcal H)}_{Fq,tF}(x,x')
&=
\frac{1}{2k_2^-k_q^+}\,
\gamma_\lambda\gamma^\eta
\left[
\frac{2k_2^r k_{2\eta}}{2k_2^-k_q^+}
+
\frac{i k_{2\eta}\sigma^{-r}}{2k_2^-}
\right]\gamma^\mu
\na[4pt]
&\approx
\frac{z}{ {{\tilde P^2}} }\,
\gamma_\lambda\gamma^\eta
\left[
-\frac{2zP_r}{ {{\tilde P^2}} }
\left(
\bar z q^-\delta_\eta^+
+\frac{P_\perp^2}{2\bar z q^-}\delta_\eta^-
+P_i\delta_\eta^i
\right)
+
\frac{i
\left(
\bar z q^-\delta_\eta^+
+\frac{P_\perp^2}{2\bar z q^-}\delta_\eta^-
+P_i\delta_\eta^i
\right)\sigma^{-r}}{2\bar z q^-}
\right]\gamma^\mu,
\na
\mathcal B^{(\mathcal H)}_{qF,Ft}(x,x')
&=
\mathcal B^{(\mathcal H)}_{Fq,Ft}(x,x')
=
\mathcal B^{(\mathcal H)}_{qF,tF}(x,x'),
\na
\mathcal B^{(\mathcal H)}_{\rm loc}
&=
-\frac{1}{2k_2^-k_q^+}\,
\gamma_\lambda\gamma^\eta\,
\delta_\eta^-\,
\frac{\sigma^{-r}}{2k_2^-}\,
\gamma^\mu
\approx
-\frac{z}{ {{\tilde P^2}} }\,
\gamma_\lambda\gamma^\eta\,
\delta_\eta^-\,
\frac{\sigma^{-r}}{2\bar z q^-}\,
\gamma^\mu =0,
\na 
\mathcal B^{(\mathcal H)}_{\rm ct}
&= - \frac{1}{2k_2^-k_q^+}\,\gamma_\lambda\,\gamma^\eta\,\left[\frac{2k_{2i} k_{qi}\,k_{2\eta}}{2k_2^- k_q^+}\,-\,\frac{2k_{2i} k_{qi} g_{\eta}^{\,\,-}}{2k_2^-}\,- g_{\eta}^i k_{qi}\right]\gamma^\mu \na
&\approx
\frac{z}{ {{\tilde P^2}} }\,
\gamma_\lambda\gamma^\eta
\left[
\frac{2z(\Delta \cdot P )}{ {{\tilde P^2}} }
\left(
\bar z q^-\delta_\eta^+
+\frac{P_\perp^2}{2\bar z q^-}\delta_\eta^-
+P_i\delta_\eta^i
\right)
-
\frac{\Delta \cdot P }{\bar z q^-}g_\eta^{\ -}
+
g_\eta^{\ i}\Delta_i
\right]\gamma^\mu
\label{eq:CH_DH_GH_KH}
\end{align}

The contribution from the external gluon line is
\begin{align}
i\mathcal M_{qg,\mathcal K}^{(a,3)}
&=
-e 
\int d^{\,4}x
\int_{x^-}^{\infty} dx^{\prime -}\,
 {e^{i k_q \cdot x } }
\epsilon_a^\lambda(k_2)\,
\epsilon_\mu(q)  
\Big[
\big(\bar\Psi(x)\,t^a\,\bar F_{-r}(x')\big)^l\,
\mathcal B^{(\mathcal K)}_{qF,tF}(x,x')
+
\big(\bar\Psi(x)\,\bar F_{-r}(x')\,t^a\big)^l\,
\mathcal B^{(\mathcal K)}_{qF,Ft}(x,x')
\Big]v^l(k_1).
\label{eq:Mqg_a_K}
\end{align}
where 
\begin{align}
\mathcal B^{(\mathcal K)}_{qF,tF}(x,x')
&=
\frac{i }{2 { {\left(k_2^-\right)}^2} k_q^+}
\left[
 k_{2r}\, g_{\nu\lambda}
\left(x^{\prime -}-x^-\right)
+
 i
\left(
\delta_\nu^- \delta_\lambda^r
-
\delta_\nu^r \delta_\lambda^-
\right)
\right]
\gamma^\nu
\left[
\left(
\frac{k_{2\perp}^2}{2k_2^-}
-
k_q^+
\right)\gamma^-
+
k_2^-\gamma^+
-
k_{2i}\gamma^i
\right]
\gamma^\mu \na [4pt]
& \approx
\frac{iz}{ \bar z\,q^- {{\tilde P^2}} }
\left[
- P_r g_{\nu\lambda}(x^{\prime -}-x^-)
+
 i\left(
\delta_\nu^-\delta_\lambda^r
-
\delta_\nu^r\delta_\lambda^-
\right)
\right]
\gamma^\nu
\left[
\left(
\frac{P_\perp^2}{2\bar z q^-}
-
\frac{ {{\tilde P^2}} }{2z\bar z q^-}
\right)\gamma^-
+
\bar z q^-\gamma^+
+
P_i\gamma^i
\right]
\gamma^\mu
\\[4pt]
\mathcal B^{(\mathcal K)}_{qF,Ft}(x,x')
&=
-
\mathcal B^{(\mathcal K)}_{qF,tF}(x,x') .
\label{eq:CK_DK_coefficients}
\end{align}

\subsection{Second quark--gluon topology}

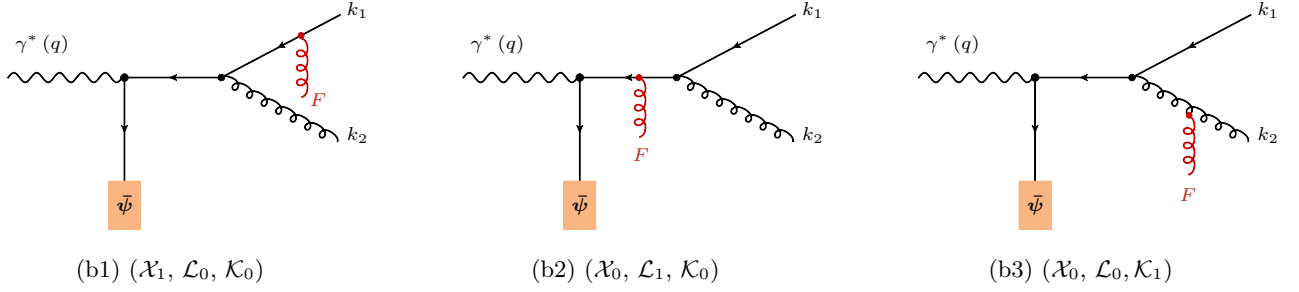
\begin{figure*}[t]
\centering
\begin{tabular}{@{}c@{\hspace{0.08\textwidth}}c@{\hspace{0.08\textwidth}}c@{}}
\begin{overpic}[
    width=0.25\textwidth
]{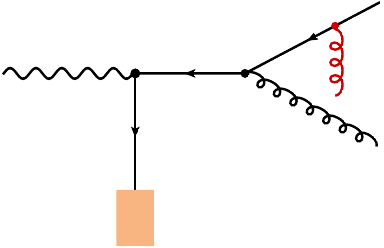}
    \put(3,54){\scriptsize $ {\gamma^*\,(q)}$}
    \put(101,64){\scriptsize $ {k_1}$}
    \put(101,27){\scriptsize $ {k_2}$}
    \put(90,37){\scriptsize $ \color{myred}{F}$}
    \put(33,6){\scriptsize $\boldsymbol{\bar\psi}$}
\end{overpic}
&
\begin{overpic}[
    width=0.25\textwidth
]{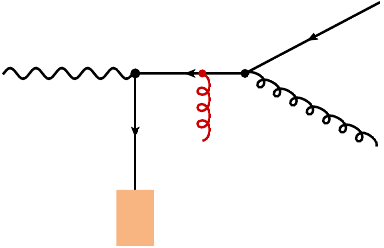}
    \put(3,54){\scriptsize $ {\gamma^*\,(q)}$}
    \put(101,64){\scriptsize $ {k_1}$}
    \put(101,27){\scriptsize $ {k_2}$}
    \put(51,21){\scriptsize $ \color{myred}{F}$}
     \put(33,6){\scriptsize $\boldsymbol{\bar\psi}$}
\end{overpic}
&
\begin{overpic}[
    width=0.25\textwidth
]{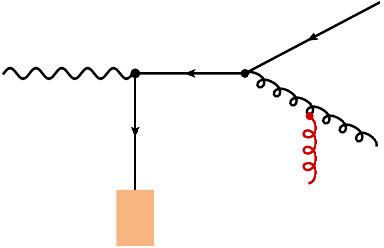}
    \put(3,54){\scriptsize $ {\gamma^*\,(q)}$}
    \put(101,64){\scriptsize $ {k_1}$}
    \put(101,27){\scriptsize $ {k_2}$}
    \put(78,9){\scriptsize $ \color{myred}{F}$}
    \put(33,6){\scriptsize $\boldsymbol{\bar\psi}$}
\end{overpic}
\\[4pt]
{\small (b1) $({\mathcal X}_1,\,{\mathcal L}_0,\,\mathcal K_0)$}
&
{\small (b2) $({\mathcal X}_0,\,{\mathcal L}_1,\,\mathcal K_0)$}
&
{\small (b3) $({\mathcal X}_0,\,{\mathcal L}_0,\mathcal K_1)$}
\end{tabular}
\caption{
Field-strength insertion points in the trilocal quark--gluon amplitude for
topology \((b)\).
}
\label{fig:qg_3body_topology_b}
\end{figure*}

The three possible field-strength insertions in topology \((b)\) are shown in Fig.~\ref{fig:qg_3body_topology_b}.  As in topology \((a)\), the insertion can be assigned to the external quark block, the internal quark-line block, or the external gluon block.
We use the shorthand notation 
\begin{align}
k_{12}^{\mu}
\equiv
k_1^\mu+k_2^\mu
\simeq
\left(
q^+ + \frac{ {{\tilde P^2}} }{2z\bar z q^-},
\ q^-,
\ \mathbf{0}_\perp
\right),
\qquad
D_{12}
\equiv
k_{12}^+
-
\frac{k_{12\perp}^2}{2k_{12}^-}
=
\frac{k_{12}^2}{2k_{12}^-}
\simeq
\frac{P_\perp^2}{2z\bar z q^-},
\label{eq:D12_def}
\end{align}
and the on-shell projected momentum
\begin{align}
p_{12}^+
=
\frac{k_{12\perp}^2}{2k_{12}^-},
\qquad
p_{12}^- = k_{12}^-,
\qquad
p_{12\perp}=k_{12\perp},
\qquad
p_{12}^\mu\simeq
\left(0,\ q^-,\ \mathbf{0}_\perp\right).
\label{eq:p12_def}
\end{align}
The relative phase between the two topology-\(b\) contributions is
\begin{align}
E(x,x')
&\equiv
\exp\left[
i\left(
p_{12}^+ - q^+
\right)x^-
+
iD_{12}x^{\prime -} -i k_q^+ x^-
\right] .
\label{eq:E12_phase_def}
\end{align}

The contribution from the LSZ-reduced external quark line is
\begin{align}
i\mathcal M_{qg,{\mathcal X}}^{(b,3)}
&=
-e  
\int d^{\,4}x
\int_{x^-}^{\infty} dx^{\prime -}\,
 {  e^{\,i k_q\cdot x}}\,
\epsilon_a^\lambda(k_2)\,
\epsilon_\mu(q)
\na
&\quad \times
\Big[ 
\big(\bar\Psi(x)\,t^a\,\bar F_{-r}(x')\big)^l
\,\mathcal D^{({\mathcal X})}_{qF,tF}(x,x')
+
E(x,x')\,
\big(\bar\Psi(x)\,t^a\,\bar F_{-r}(x')\big)^l
\,\mathcal D^{({\mathcal X},E)}_{qF,tF}(x,x')
\Big]v^l(k_1).
\label{eq:Mqg_b_A}
\end{align}
where 
\begin{align}
\mathcal D^{({\mathcal X})}_{qF,tF}(x,x')
&=
\frac{i }{4 k_1^- k_{12}^- D_{12}}\,
\gamma^\mu\gamma^\nu\gamma_\lambda
\Bigg[
2k_{1r}k_{12,\nu}(x^{\prime -}-x^-)
-
k_{12,\nu}\sigma^{-r}
+
\frac{2i\,p_{12,\nu}k_{1r}}{D_{12}}
\Bigg]\na [4pt]
& \approx
\frac{i\bar z}{2q^-P_\perp^2}\,
\gamma^\mu\gamma^\nu\gamma_\lambda
\Bigg[
2P_r\left(q^-\delta_\nu^+ + \frac{P_\perp^2}{2z\bar z q^-}\delta_\nu^-\right)(x^{\prime -}-x^-)
-
\left(q^-\delta_\nu^+ + \frac{P_\perp^2}{2z\bar z q^-}\delta_\nu^-\right)\sigma^{-r}
+
\frac{4iz\bar z q^-}{P_\perp^2}
q^-\delta_\nu^+ P_r
\Bigg]
\\[4pt]
\mathcal D^{({\mathcal X},E)}_{qF,tF}(x,x')
&=
\frac{i  }{{ 4 k_1^-}k_{12}^-D_{12}}\,
\gamma^\mu\gamma^\nu\gamma_\lambda
\Bigg[
-
\frac{2i\,p_{12,\nu}k_{1r}}{D_{12}}
-
p_{12,\nu}\sigma^{-r}
\Bigg]\na [4pt]
& \approx
\frac{i\bar z}{2 q^- P_\perp^2}\,
\gamma^\mu\gamma^\nu\gamma_\lambda
\Bigg[
-
\frac{4iz\bar z q^-}{P_\perp^2}
q^-\delta_\nu^+ P_r
-
q^-\delta_\nu^+\sigma^{-r}
\Bigg]
\label{eq:CA_b_coefficients}
\end{align}

The contribution from the subleading term in the internal quark
propagator is
\begin{align}
i\mathcal M_{qg,\mathcal  L}^{(b,3)}
&=
-e  {\, \frac{1}{(2q^-)}}
\int d^{\,4}x
\int_{x^-}^{\infty} dx^{\prime -}\,
 { e^{\,i k_q\cdot x }}\,
\epsilon_a^\lambda(k_2)\,
\epsilon_\mu(q)
\na
&\quad \times
\Big[ 
\big(\bar\Psi(x)\,\bar F_{-r}(x')\,t^a\big)^l
\,\mathcal D^{(\mathcal L)}_{qF,Ft}(x,x')
+
E(x,x')\,
\big(\bar\Psi(x)\,\bar F_{-r}(x')\,t^a\big)^l
\,\mathcal D^{(\mathcal L, E)}_{qF,Ft}(x,x')
\Big]v^l(k_1).
\label{eq:Mqg_b_L}
\end{align}
where 
\begin{align}
{ \mathcal D^{(\mathcal L)}_{qF,Ft}(x,x')}
&=
\frac{1}{4(k_{12}^-)^2D_{12}^2}\,
\gamma^\mu\gamma^\beta
\Bigg[
2p_{12,r}p_{12,\beta}
+
2p_{12}^-D_{12}
\left(
\frac{p_{12,r}}{p_{12}^-}\delta_\beta^-
+
\delta_\beta^r
\right)
\Bigg]\gamma_\lambda\na [4pt]
& \approx
\frac{z\bar z}{ P_\perp^2}\,
\gamma^\mu\gamma^\beta
\left(
\delta_\beta^r
\right)
 \gamma_\lambda
\\[4pt]
{ \mathcal D^{(\mathcal L, E)}_{qF,Ft}(x,x')}
&=
{ \frac{i}{4(k_{12}^-)^2D_{12}^2}\,
\gamma^\mu
\Bigg[
2p_{12,r}\slashed p_{12}(x^{\prime -}-x^-)
+i\left(
2p_{12,r}\slashed p_{12}
+
2p_{12}^-D_{12}
\right)
\left(
\frac{p_{12,r}}{p_{12}^-}\gamma^-
+
\gamma^r
\right)
+
D_{12}\sigma^{-r}\gamma^-}
\na
&\qquad
-
2p_{12,r}\slashed k_{12}(x^{\prime -}-x^-)
-
\sigma^{-r}\slashed k_{12}
\Bigg]\gamma_\lambda \na [4pt]
& \approx
\frac{iz^2\bar z^2}{ P_\perp^4}\,
\gamma^\mu
\Bigg[
i\left(
\frac{P_\perp^2}{z\bar z}
\right)
\left(
\gamma^r
\right)
+
\frac{P_\perp^2}{2z\bar z q^-}\sigma^{-r}\gamma^-
-
\sigma^{-r}\left[\left(q^+ + \frac{ {{\tilde P^2}} }{2z\bar z q^-}\right)\gamma^- + q^-\gamma^+\right]
\Bigg]\gamma_\lambda
\label{eq:DL_b_coefficients}
\end{align}

The contribution from the external gluon line is
\begin{align}
i\mathcal M_{qg,\mathcal K}^{(b,3)}
&=
-e  {\, \frac{1}{(2q^-)}}
\int d^{\,4}x
\int_{x^-}^{\infty} dx^{\prime -}\,
 { e^{\,i k_q\cdot x }}\,
\epsilon_a^\lambda(k_2)\,
\epsilon_\mu(q)
\na
&\quad \times
\Bigg[ 
\Big(
\big(\bar\Psi(x)\,t^a\,\bar F_{-r}(x')\big)^l
\,\mathcal D^{(\mathcal K)}_{qF,tF}(x,x')
+
\big(\bar\Psi(x)\,\bar F_{-r}(x')\,t^a\big)^l
\,\mathcal D^{(\mathcal K)}_{qF,Ft}(x,x')
\Big)
\na
&\qquad
+
E(x,x')
\Big(
\big(\bar\Psi(x)\,t^a\,\bar F_{-r}(x')\big)^l
\,\mathcal D^{(\mathcal K,E)}_{qF,tF}(x,x')
+
\big(\bar\Psi(x)\,\bar F_{-r}(x')\,t^a\big)^l
\,\mathcal D^{(\mathcal K,E)}_{qF,Ft}(x,x')
\Big)
\Bigg]v^l(k_1).
\label{eq:Mqg_b_K}
\end{align}
where
\begin{align}
\mathcal D^{(\mathcal K)}_{qF,tF}(x,x')
&=
-\frac{i }{4 k_2^- k_{12}^-D_{12}}\,
\gamma^\mu\gamma^\beta\gamma^\nu
\Bigg[
2g_{\nu\lambda}k_{12,\beta}k_{2r}(x^{\prime -}-x^-)
+
2i k_{12,\beta}
\left(
\delta_\nu^-\delta_\lambda^r
-
\delta_\nu^r\delta_\lambda^-
\right)
+
\frac{2i\,g_{\nu\lambda}p_{12,\beta}k_{2r}}{D_{12}}
\Bigg]\na [4pt]
& \approx
-\frac{iz}{2q^-P_\perp^2}\,
\gamma^\mu\gamma^\beta\gamma^\nu
\Bigg[
2g_{\nu\lambda}
\left(
q^-\delta_\beta^+
+
\frac{P_\perp^2}{2z\bar z q^-}\delta_\beta^-
\right)(-P_r)
(x^{\prime -}-x^-)
+
2i
\left(
q^-\delta_\beta^+
+
\frac{P_\perp^2}{2z\bar z q^-}\delta_\beta^-
\right)
\left(
\delta_\nu^-\delta_\lambda^r
-
\delta_\nu^r\delta_\lambda^-
\right)
\na
&\qquad
+
\frac{4iz\bar z q^-}{P_\perp^2}
g_{\nu\lambda}q^-\delta_\beta^+(-P_r)
\Bigg]
\\[4pt]
\mathcal D^{(\mathcal K)}_{qF,Ft}(x,x')
&=
-
\mathcal D^{(\mathcal K)}_{qF,tF}(x,x'),
\\[4pt]
\mathcal D^{(\mathcal K,E)}_{qF,tF}(x,x')
&=
-\frac{1}{4 k_2^- k_{12}^-D_{12}}\,
\gamma^\mu\gamma^\beta\gamma^\nu
\Bigg[
\frac{2g_{\nu\lambda}p_{12,\beta}k_{2r}}{D_{12}}
+
2p_{12,\beta}
\left(
\delta_\nu^-\delta_\lambda^r
-
\delta_\nu^r\delta_\lambda^-
\right)
\Bigg]\na [4pt]
& \approx
-\frac{z}{2q^-P_\perp^2}\,
\gamma^\mu\gamma^\beta\gamma^\nu
\Bigg[
\frac{4z\bar z q^-}{P_\perp^2}
g_{\nu\lambda}q^-\delta_\beta^+(-P_r)
+
2q^-\delta_\beta^+
\left(
\delta_\nu^-\delta_\lambda^r
-
\delta_\nu^r\delta_\lambda^-
\right)
\Bigg]
\\[4pt]
\mathcal D^{(\mathcal K,E)}_{qF,Ft}(x,x')
&=
-
\mathcal D^{(\mathcal K,E)}_{qF,tF}(x,x') .
\label{eq:CK_b_coefficients}
\end{align}

\subsection{Summed trilocal coefficient structures}

We now collect all the elements into total coefficients multiplying identical operator, phase, and Dirac structures. For the first topology the sum of Eqs.~\eqref{eq:Mqg_a_A}, \eqref{eq:Mqg_a_H}, and \eqref{eq:Mqg_a_K} can be written as
\begin{align}
i\mathcal M_{qg,\bar\psi}^{(a,3)}
&=-e  
\int d^{\,4}x
\int_{x^-}^{\infty} dx^{\prime -}\,
e^{i k_q\cdot x}\,
\epsilon_a^\lambda(k_2)\,
\epsilon_\mu(q)\,
\mathcal C_{(a)}^{\lambda\mu}(x,x')\,v(k_1)\na
&=
-e  
\int d^{\,4}x\,
e^{i k_q\cdot x}\,
\epsilon_a^\lambda(k_2)\,
\epsilon_\mu(q)
\Bigg\{
\int_{x^-}^{\infty} dx^{\prime -}
\Big[
\big(\bar\Psi(x)t^a\bar F_{-r}(x')\big)^l\,
\mathcal C^{(a)}_{qF,tF}(x,x')
+
\big(\bar\Psi(x)\bar F_{-r}(x')t^a\big)^l\,
\mathcal C^{(a)}_{qF,Ft}(x,x')\na
&\qquad
+
\big(\bar\Psi(x')t^a\bar F_{-r}(x)\big)^l\,
\mathcal C^{(a)}_{Fq,tF}(x,x')
+
\big(\bar\Psi(x')\bar F_{-r}(x)t^a\big)^l\,
\mathcal C^{(a)}_{Fq,Ft}(x,x')
\Big]
+
\big(\bar\Psi(x)t^a\bar F_{-r}(x)\big)^l\,
\mathcal C^{(a)}_{\rm loc}
+
\big(\bar\Psi(x)t^a\big)^l\,
\mathcal C^{(a)}_{\rm ct}
\Bigg\}
v^l(k_1),
\label{eq:Mqg_a_summed}
\end{align}
with
\begin{align}
\mathcal C^{(a)}_{qF,tF}(x,x')
&\equiv
\mathcal B^{({\mathcal X})}_{qF,tF}(x,x')
+
\mathcal B^{(\mathcal H)}_{qF,tF}(x,x')
+
\mathcal B^{(\mathcal K)}_{qF,tF}(x,x'),
\na [3pt]
\mathcal C^{(a)}_{qF,Ft}(x,x')
&\equiv
\mathcal B^{(\mathcal H)}_{qF,Ft}(x,x')
+
\mathcal B^{(\mathcal K)}_{qF,Ft}(x,x')
=
\mathcal B^{(\mathcal H)}_{qF,Ft}(x,x')
-
\mathcal B^{(\mathcal K)}_{qF,tF}(x,x'),
\na [3pt]
\mathcal C^{(a)}_{Fq,tF}(x,x')
&\equiv
\mathcal B^{({\mathcal X})}_{Fq,tF}(x,x')
+
\mathcal B^{(\mathcal H)}_{Fq,tF}(x,x'),
\na [3pt]
\mathcal C^{(a)}_{Fq,Ft}(x,x')
&\equiv
\mathcal B^{(\mathcal H)}_{Fq,Ft}(x,x'),\qquad
\mathcal C^{(a)}_{\rm loc}
\equiv
\mathcal B^{(\mathcal H)}_{\rm loc},
\qquad
\mathcal C^{(a)}_{\rm ct}
\equiv
\mathcal B^{(\mathcal H)}_{\rm ct}.
\label{eq:Ca_summed_coefficients}
\end{align}
Thus all nonlocal first-topology contributions are reduced to four coefficient structures before the color contraction: the two position orderings \(qF\) and \(Fq\), each with the relevant \(tF\) or \(Ft\) color ordering.

For the second topology the common phase and color structures can be combined in the same way.  The full result following from Eqs.~\eqref{eq:Mqg_b_A}, \eqref{eq:Mqg_b_L}, and \eqref{eq:Mqg_b_K} is
\begin{align}
i\mathcal M_{qg,\bar\psi}^{(b,3)}
&=-e  
\int d^{\,4}x
\int_{x^-}^{\infty} dx^{\prime -}\,
e^{\,i k_q \cdot x}\,
\epsilon_a^\lambda(k_2)\,
\epsilon_\mu(q)\,
\mathcal C_{(b)}^{\lambda\mu}(x,x')\,v(k_1)\na[4pt]
&=
-e 
\int d^{\,4}x
\int_{x^-}^{\infty} dx^{\prime -}\,
e^{\,i k_q \cdot x}\,
\epsilon_a^\lambda(k_2)\,
\epsilon_\mu(q)
\na
&\quad\times
\Bigg[
\big(\bar\Psi(x)t^a\bar F_{-r}(x')\big)^l\,
\mathcal C^{(b)}_{qF,tF}(x,x')
+
\big(\bar\Psi(x)\bar F_{-r}(x')t^a\big)^l\,
\mathcal C^{(b)}_{qF,Ft}(x,x')
\na
&\qquad
+
E(x,x')
\Big(
\big(\bar\Psi(x)t^a\bar F_{-r}(x')\big)^l\,
\mathcal C^{(b,E)}_{qF,tF}(x,x')
+
\big(\bar\Psi(x)\bar F_{-r}(x')t^a\big)^l\,
\mathcal C^{(b,E)}_{qF,Ft}(x,x')
\Big)
\Bigg]v^l(k_1),
\label{eq:Mqg_b_summed}
\end{align}
where
\begin{align}
\mathcal C^{(b)}_{qF,tF}(x,x')
&\equiv
\mathcal D^{({\mathcal X})}_{qF,tF}(x,x')
+
\mathcal D^{(\mathcal K)}_{qF,tF}(x,x'),
\na [3pt]
\mathcal C^{(b)}_{qF,Ft}(x,x')
&\equiv
\mathcal D^{(\mathcal L)}_{qF,Ft}(x,x')
+
\mathcal D^{(\mathcal K)}_{qF,Ft}(x,x')
=
\mathcal D^{(\mathcal L)}_{qF,Ft}(x,x')
-
\mathcal D^{(\mathcal K)}_{qF,tF}(x,x'),
\na [3pt]
\mathcal C^{(b,E)}_{qF,tF}(x,x')
&\equiv
\mathcal D^{({\mathcal X},E)}_{qF,tF}(x,x')
+
\mathcal D^{(\mathcal K,E)}_{qF,tF}(x,x'),
\na [3pt]
\mathcal C^{(b,E)}_{qF,Ft}(x,x')
&\equiv
\mathcal D^{(\mathcal L, E)}_{qF,Ft}(x,x')
+
\mathcal D^{(\mathcal K,E)}_{qF,Ft}(x,x')
=
\mathcal D^{(\mathcal L, E)}_{qF,Ft}(x,x')
-
\mathcal D^{(\mathcal K,E)}_{qF,tF}(x,x') .
\label{eq:Cb_summed_coefficients}
\end{align}
Equations~\eqref{eq:Ca_summed_coefficients} and \eqref{eq:Cb_summed_coefficients} are the desired summed coefficients: each \(\mathcal C\) absorbs all contributions with the same position ordering, color ordering, and longitudinal phase structure.

\bibliography{quark_dijet}

@article{Altinoluk:2023qfr,
    author = "Altinoluk, Tolga and Armesto, Nestor and Beuf, Guillaume",
    title = "{Probing quark transverse momentum distributions in the color glass condensate: Quark-gluon dijets in deep inelastic scattering at next-to-eikonal accuracy}",
    eprint = "2303.12691",
    archivePrefix = "arXiv",
    primaryClass = "hep-ph",
    doi = "10.1103/PhysRevD.108.074023",
    journal = "Phys. Rev. D",
    volume = "108",
    number = "7",
    pages = "074023",
    year = "2023"
}

@article{Ji:2004wu,
    author = "Ji, Xiang-dong and Ma, Jian-ping and Yuan, Feng",
    title = "{QCD factorization for semi-inclusive deep-inelastic scattering at low transverse momentum}",
    eprint = "hep-ph/0404183",
    archivePrefix = "arXiv",
    reportNumber = "DOE-ER-40762-308, UM-PP-04-037, UM-PP-{\#}04-037",
    doi = "10.1103/PhysRevD.71.034005",
    journal = "Phys. Rev. D",
    volume = "71",
    pages = "034005",
    year = "2005"
}

@book{Collins:2011zzd,
    author = "Collins, John",
    title = "{Foundations of Perturbative QCD}",
    doi = "10.1017/9781009401845",
    isbn = "978-1-009-40184-5, 978-1-009-40183-8, 978-1-009-40182-1",
    publisher = "Cambridge University Press",
    volume = "32",
    year = "2011"
}

@article{Boussarie:2023izj,
    author = "Boussarie, Renaud and others",
    title = "{TMD Handbook}",
    eprint = "2304.03302",
    archivePrefix = "arXiv",
    primaryClass = "hep-ph",
    reportNumber = "JLAB-THY-23-3780, LA-UR-21-20798, MIT-CTP/5386",
    journal = "arXiv e-prints",
    month = "4",
    year = "2023"
}

@article{Mukherjee:2026cte,
    author = "Mukherjee, Swagato and Skokov, Vladimir V. and Tarasov, Andrey and Tiwari, Shaswat and Yao, Fei",
    title = "{Back-to-back dijet production in DIS at arbitrary Bjorken-x: TMD gluon distributions to twist-3 accuracy}",
    eprint = "2602.15137",
    archivePrefix = "arXiv",
    primaryClass = "hep-ph",
    journal = "arXiv e-prints",
    month = "2",
    year = "2026"
}

@article{Accardi:2012qut,
    author = "Accardi, A. and others",
    title = "{Electron Ion Collider: The Next QCD Frontier: Understanding the glue that binds us all}",
    eprint = "1212.1701",
    archivePrefix = "arXiv",
    primaryClass = "nucl-ex",
    doi = "10.1140/epja/i2016-16268-9",
    journal = "Eur. Phys. J. A",
    volume = "52",
    number = "9",
    pages = "268",
    year = "2016"
}

@article{Dominguez:2011br,
    author = "Dominguez, Fabio and Qiu, Jian-Wei and Xiao, Bo-Wen and Yuan, Feng",
    title = "{On the linearly polarized gluon distributions in the color dipole model}",
    eprint = "1109.6293",
    archivePrefix = "arXiv",
    primaryClass = "hep-ph",
    doi = "10.1103/PhysRevD.85.045003",
    journal = "Phys. Rev. D",
    volume = "85",
    pages = "045003",
    year = "2012"
}

@article{Metz:2011wb,
    author = "Metz, Andreas and Zhou, Jian",
    title = "{Distribution of linearly polarized gluons inside a large nucleus}",
    eprint = "1105.1991",
    archivePrefix = "arXiv",
    primaryClass = "hep-ph",
    doi = "10.1103/PhysRevD.84.051503",
    journal = "Phys. Rev. D",
    volume = "84",
    pages = "051503",
    year = "2011"
}

@article{AbdulKhalek:2021gbh,
    author = "Abdul Khalek, R. and others",
    title = "{Science Requirements and Detector Concepts for the Electron-Ion Collider: EIC Yellow Report}",
    eprint = "2103.05419",
    archivePrefix = "arXiv",
    primaryClass = "physics.ins-det",
    doi = "10.1016/j.nuclphysa.2022.122447",
    journal = "Nucl. Phys. A",
    volume = "1026",
    pages = "122447",
    year = "2022"
}

@article{McLerran:1994vd,
    author = "McLerran, Larry D. and Venugopalan, Raju",
    title = "{Green's functions in the color field of a large nucleus}",
    eprint = "hep-ph/9402335",
    archivePrefix = "arXiv",
    reportNumber = "TPI-MINN-94-7-T, NUC-MINN-94-2-T, HEP-MINN-94-1242-T",
    doi = "10.1103/PhysRevD.50.2225",
    journal = "Phys. Rev. D",
    volume = "50",
    pages = "2225--2233",
    year = "1994"
}

@inbook{Iancu:2003xm,
    author = "Iancu, Edmond and Venugopalan, Raju",
    editor = "Hwa, Rudolph C. and Wang, Xin-Nian",
    title = "{The Color glass condensate and high-energy scattering in QCD}",
    booktitle = "{Quark-gluon plasma 4}",
    eprint = "hep-ph/0303204",
    archivePrefix = "arXiv",
    doi = "10.1142/9789812795533_0005",
    pages = "249--3363",
    month = "3",
    year = "2003"
}

@article{Dominguez:2011wm,
    author = "Dominguez, Fabio and Marquet, Cyrille and Xiao, Bo-Wen and Yuan, Feng",
    title = "{Universality of Unintegrated Gluon Distributions at small x}",
    eprint = "1101.0715",
    archivePrefix = "arXiv",
    primaryClass = "hep-ph",
    doi = "10.1103/PhysRevD.83.105005",
    journal = "Phys. Rev. D",
    volume = "83",
    pages = "105005",
    year = "2011"
}

@article{McLerran:1993ni,
    author = "McLerran, Larry D. and Venugopalan, Raju",
    title = "{Computing quark and gluon distribution functions for very large nuclei}",
    eprint = "hep-ph/9309289",
    archivePrefix = "arXiv",
    doi = "10.1103/PhysRevD.49.2233",
    journal = "Phys. Rev. D",
    volume = "49",
    pages = "2233--2241",
    year = "1994"
}

@article{Gelis:2010nm,
    author = "Gelis, Francois and Iancu, Edmond and Jalilian-Marian, Jamal and Venugopalan, Raju",
    title = "{The Color Glass Condensate}",
    eprint = "1002.0333",
    archivePrefix = "arXiv",
    primaryClass = "hep-ph",
    doi = "10.1146/annurev.nucl.010909.083629",
    journal = "Ann. Rev. Nucl. Part. Sci.",
    volume = "60",
    pages = "463--489",
    year = "2010"
}

@article{Mulders:2000sh,
    author = "Mulders, P. J. and Rodrigues, J.",
    title = "{Transverse momentum dependence in gluon distribution and fragmentation functions}",
    eprint = "hep-ph/0009343",
    archivePrefix = "arXiv",
    reportNumber = "VUTH-00-23",
    doi = "10.1103/PhysRevD.63.094021",
    journal = "Phys. Rev. D",
    volume = "63",
    pages = "094021",
    year = "2001"
}

@article{delCastillo:2020omr,
    author = "del Castillo, Rafael F. and Echevarria, Miguel G. and Makris, Yiannis and Scimemi, Ignazio",
    title = "{TMD factorization for dijet and heavy-meson pair in DIS}",
    eprint = "2008.07531",
    archivePrefix = "arXiv",
    primaryClass = "hep-ph",
    doi = "10.1007/JHEP01(2021)088",
    journal = "JHEP",
    volume = "01",
    pages = "088",
    year = "2021"
}

@article{Mukherjee:2023snp,
    author = "Mukherjee, Swagato and Skokov, Vladimir V. and Tarasov, Andrey and Tiwari, Shaswat",
    title = "{Unified description of DGLAP, CSS, and BFKL evolution: TMD factorization bridging large and small x}",
    eprint = "2311.16402",
    archivePrefix = "arXiv",
    primaryClass = "hep-ph",
    doi = "10.1103/PhysRevD.109.034035",
    journal = "Phys. Rev. D",
    volume = "109",
    number = "3",
    pages = "034035",
    year = "2024"
}

@article{Mukherjee:2025aiw,
    author = "Mukherjee, Swagato and Skokov, Vladimir V. and Tarasov, Andrey and Tiwari, Shaswat",
    title = "{Perturbative corrections to quark TMDPDFs in the background-field method: Gauge invariance, equations of motion, and multiple interactions}",
    eprint = "2502.15889",
    archivePrefix = "arXiv",
    primaryClass = "hep-ph",
    doi = "10.1103/zrz3-3rbh",
    journal = "Phys. Rev. D",
    volume = "111",
    number = "11",
    pages = "114034",
    year = "2025"
}

@article{Singh:2026wol,
    author = "Singh, Siddharth Narayan and Aggarwal, R. and Kaur, M.",
    title = "{Study of quark and gluon jet identification in photoproduction at EIC}",
    eprint = "2603.09545",
    archivePrefix = "arXiv",
    primaryClass = "hep-ex",
    journal = "",
    month = "3",
    year = "2026"
}

@article{Gras:2017jty,
    author = {Gras, Philippe and H{\"o}che, Stefan and Kar, Deepak and Larkoski, Andrew and L{\"o}nnblad, Leif and Pl{\"a}tzer, Simon and Si{\'o}dmok, Andrzej and Skands, Peter and Soyez, Gregory and Thaler, Jesse},
    title = "{Systematics of quark/gluon tagging}",
    eprint = "1704.03878",
    archivePrefix = "arXiv",
    primaryClass = "hep-ph",
    reportNumber = "MIT-CTP-4885, COEPP-MN-17-2, MCNET-17-04",
    doi = "10.1007/JHEP07(2017)091",
    journal = "JHEP",
    volume = "07",
    pages = "091",
    year = "2017"
}

@article{Frye:2017yrw,
    author = "Frye, Christopher and Larkoski, Andrew J. and Thaler, Jesse and Zhou, Kevin",
    title = "{Casimir Meets Poisson: Improved Quark/Gluon Discrimination with Counting Observables}",
    eprint = "1704.06266",
    archivePrefix = "arXiv",
    primaryClass = "hep-ph",
    reportNumber = "MIT-CTP-4987, MIT--CTP-4987",
    doi = "10.1007/JHEP09(2017)083",
    journal = "JHEP",
    volume = "09",
    pages = "083",
    year = "2017"
}

@article{Boussarie:2021ybe,
    author = {Boussarie, Renaud and M{\"a}ntysaari, Heikki and Salazar, Farid and Schenke, Bj{\"o}rn},
    title = "{The importance of kinematic twists and genuine saturation effects in dijet production at the Electron-Ion Collider}",
    eprint = "2106.11301",
    archivePrefix = "arXiv",
    primaryClass = "hep-ph",
    doi = "10.1007/JHEP09(2021)178",
    journal = "JHEP",
    volume = "09",
    pages = "178",
    year = "2021"
}

@article{Caucal:2023fsf,
    author = {Caucal, Paul and Salazar, Farid and Schenke, Bj{\"o}rn and Stebel, Tomasz and Venugopalan, Raju},
    title = "{Back-to-Back Inclusive Dijets in Deep Inelastic Scattering at Small x: Complete NLO Results and Predictions}",
    eprint = "2308.00022",
    archivePrefix = "arXiv",
    primaryClass = "hep-ph",
    doi = "10.1103/PhysRevLett.132.081902",
    journal = "Phys. Rev. Lett.",
    volume = "132",
    number = "8",
    pages = "081902",
    year = "2024"
}

@article{Caucal:2021ent,
    author = "Caucal, Paul and Salazar, Farid and Venugopalan, Raju",
    title = "{Dijet impact factor in DIS at next-to-leading order in the Color Glass Condensate}",
    eprint = "2108.06347",
    archivePrefix = "arXiv",
    primaryClass = "hep-ph",
    doi = "10.1007/JHEP11(2021)222",
    journal = "JHEP",
    volume = "11",
    pages = "222",
    year = "2021"
}

@article{Caucal:2023nci,
    author = {Caucal, Paul and Salazar, Farid and Schenke, Bj{\"o}rn and Stebel, Tomasz and Venugopalan, Raju},
    title = {{Back-to-back inclusive dijets in DIS at small x: gluon Weizs{\"a}cker-Williams distribution at NLO}},
    eprint = "2304.03304",
    archivePrefix = "arXiv",
    primaryClass = "hep-ph",
    doi = "10.1007/JHEP08(2023)062",
    journal = "JHEP",
    volume = "08",
    pages = "062",
    year = "2023"
}

@article{Altinoluk:2022jkk,
    author = "Altinoluk, Tolga and Beuf, Guillaume and Czajka, Alina and Tymowska, Arantxa",
    title = "{DIS dijet production at next-to-eikonal accuracy in the CGC}",
    eprint = "2212.10484",
    archivePrefix = "arXiv",
    primaryClass = "hep-ph",
    doi = "10.1103/PhysRevD.107.074016",
    journal = "Phys. Rev. D",
    volume = "107",
    number = "7",
    pages = "074016",
    year = "2023"
}

@article{Altinoluk:2024zom,
    author = "Altinoluk, Tolga and Beuf, Guillaume and Czajka, Alina and Marquet, Cyrille",
    title = "{Back-to-back dijet production in DIS at next-to-eikonal accuracy and twist-3 gluon TMDs}",
    eprint = "2410.00612",
    archivePrefix = "arXiv",
    primaryClass = "hep-ph",
    doi = "10.1103/PhysRevD.111.014010",
    journal = "Phys. Rev. D",
    volume = "111",
    number = "1",
    pages = "014010",
    year = "2025"
}

@article{Dumitru:2018kuw,
    author = "Dumitru, Adrian and Skokov, Vladimir and Ullrich, Thomas",
    title = {{Measuring the Weizs{\"a}cker-Williams distribution of linearly polarized gluons at an electron-ion collider through dijet azimuthal asymmetries}},
    eprint = "1809.02615",
    archivePrefix = "arXiv",
    primaryClass = "hep-ph",
    doi = "10.1103/PhysRevC.99.015204",
    journal = "Phys. Rev. C",
    volume = "99",
    number = "1",
    pages = "015204",
    year = "2019"
}

@article{Bacchetta:2006tn,
  author        = "Bacchetta, Alessandro and Diehl, Markus and Goeke, Klaus and Metz, Andreas and Mulders, Piet J. and Schlegel, Marc",
  title         = "{Semi-inclusive deep inelastic scattering at small transverse momentum}",
  journal       = "JHEP",
  volume        = "02",
  pages         = "093",
  year          = "2007",
  doi           = "10.1088/1126-6708/2007/02/093",
  eprint        = "hep-ph/0611265",
  archivePrefix = "arXiv",
  primaryClass  = "hep-ph"
}
\end{document}